\documentclass[a4paper,number,5p,twocolumn]{elsarticle}

\usepackage[english]{babel}
\usepackage[utf8]{inputenc}
\usepackage{setspace}
\usepackage{indentfirst}
\usepackage[usenames,dvipsnames]{xcolor}
\usepackage[unicode,breaklinks=true]{hyperref}
\usepackage{graphicx}
\usepackage{float}
\usepackage{array}
\usepackage{multirow}
\usepackage{amsfonts}
\usepackage{amsmath}
\usepackage{amssymb}
\usepackage[mathscr]{euscript}
\usepackage{bm}
\usepackage{soul}
\usepackage{lineno}
\usepackage{csquotes}
\usepackage{scalerel}	
\usepackage{mathtools}
\usepackage{paralist}
\usepackage{xifthen}
\usepackage{natbib}
\usepackage{interval}

\soulregister\cite7
\soulregister\ref7
\soulregister\pageref7

\makeatletter
\providecommand{\doi}[1]{%
	\begingroup
	\let\bibinfo\@secondoftwo
	\urlstyle{rm}%
	\href{http://dx.doi.org/#1}{%
		doi:\discretionary{}{}{}%
		\nolinkurl{#1}%
	}%
	\endgroup
}
\makeatother

\DeclareGraphicsExtensions{.pdf,.eps,.png,.jpg}	

\newcommand{\Fref}[1]{Fig.~\ref{#1}}
	
\newcommand{\Eref}[1]{Eq.~(\ref{#1})}

\newcommand{\Sref}[1]{{Section~\ref{#1}}}

\newcommand{\Matlab}{MATLAB\textsuperscript{\textregistered}}

\newcommand{\tiledomg}{\ensuremath{\mathcal{T}}}				
\newcommand{\setdom}{\ensuremath{\mathcal{S}}}

\newcommand{\tiledom}[1]{\ensuremath{\mathcal{T}\ifthenelse{\isempty{#1}}{}{_{(#1)}} }}
\newcommand{\conog}[1]{\ensuremath{\mathcal{G}\ifthenelse{\isempty{#1}}{}{_{(#1)}} }}		

\newcommand{\bmath}[1]{\ensuremath{\bm{#1}}}

\mathchardef\mhyphen="2D

\DeclareMathOperator*{\assembly}{\scalerel*{\mathrm{A}}{\sum}}

\newcommand{\de}[1]{\,{\mathrm d}#1}
\newcommand{\at}[1]{\vert_{#1}}
\newcommand{\atx}{\ensuremath{(\tens{x})}}
\newcommand{\half}{\frac{1}{2}}

\newcommand{\set}[1]{{\mathbb #1}}

\newcommand{\setR}{\set{R}}

\newcommand{\domain}{\Omega}
\newcommand{\boundary}{\partial\domain}

\newcommand{\scal}[1]{\mathnormal{#1}}
\newcommand{\tens}[1]{\boldsymbol{#1}}				
\newcommand{\tenss}[1]{\bmath{#1}} 					    

\newcommand{\x}{\tens{x}} 								
\newcommand{\trn}{^{\sf T}}

\newcommand{\semtrx}[1]{\mathsf{#1}} 					
\newcommand{\sevek}[1]{\mathsf{#1}} 					

\newcommand{\scontr}{\cdot}

\newcommand{\grad}{\bm{\nabla}}
\newcommand{\divergence}{\grad \scontr}

\newcommand{\temperature}{\ensuremath{\theta}}
\newcommand{\temp}{\temperature}

\newcommand{\tFluct}{\ensuremath{\tilde{\scal{\temperature}}}}

\newcommand{\tGrad}[1]{\ensuremath{\tenss{G}^{(#1)}\!}}
\newcommand{\tGradF}{\tGrad{1}}
\newcommand{\tGradS}{\tGrad{2}}

\newcommand{\flux}{\ensuremath{\tens{q}}}
\newcommand{\vertexlabel}[1]{\large{\textbf{\textsf{#1}}}}

\setcitestyle{square,sort&compress,numbers}

\begin{document}
	
	
\begin{frontmatter}
		
\title{Microstructure-informed reduced modes synthesized with Wang tiles and the Generalized Finite Element Method}

\author[ctu]{Martin Do\v{s}k\'{a}\v{r}}
\ead{martin.doskar@fsv.cvut.cz}
\author[ctu]{Jan Zeman}
\ead{jan.zeman@cvut.cz}
\author[ucsd]{Petr Krysl}
\ead{pkrysl@ucsd.edu}
\author[ctu,unilu]{Jan Nov\'{a}k}
\ead{novakja@fsv.cvut.cz}
\address[ctu]{Faculty of Civil Engineering, Czech Technical University in Prague, Th\'{a}kurova 2077/7, \mbox{166 29 Prague 6}, Czech Republic}
\address[ucsd]{Jacobs School of Engineering, University of California, San Diego, 9500 Gilman Dr., La Jolla, CA 92093, USA}
\address[unilu]{Institute of Computational Engineering\\University of Luxembourg, Avenue de la Fonte 6, L-4364 Esch-sur-Alzette, Luxembourg }

\journal{arXiv.org}

\begin{abstract}

	A recently introduced representation by a set of Wang tiles---a generalization of the traditional Periodic Unit Cell based approach---serves as a reduced geometrical model for materials with stochastic heterogeneous microstructure, enabling an efficient synthesis of microstructural realizations.
	To facilitate macroscopic analyses with a fully resolved microstructure generated with Wang tiles, we develop a reduced order modelling scheme utilizing pre-computed characteristic features of the tiles. 
	
	In the offline phase, inspired by the computational homogenization, we extract continuous fluctuation fields from the compressed microstructural representation as responses to generalized loading represented by the first- and second-order macroscopic gradients.
	In the online phase, using the ansatz of the Generalized Finite Element Method, we combine these fields with a coarse finite element discretization to create microstructure-informed reduced modes specific for a given macroscopic problem. 
	
	Considering a two-dimensional scalar elliptic problem, we demonstrate that our scheme delivers less than a 3\% error in both the relative $L_2$ and energy norms with only 0.01\% of the unknowns when compared to the fully resolved problem. 
	Accuracy can be further improved by locally refining the macroscopic discretization and/or employing more pre-computed fluctuation fields. 
	Finally, unlike the standard snapshot-based reduced-order approaches, our scheme handles significant changes in the macroscopic geometry or loading without the need for recalculating the offline phase, because the fluctuation fields are extracted without any prior knowledge on the macroscopic problem. 
	
\end{abstract}

\begin{keyword}
	Wang tiling, Microstructure-informed modes, Reduced order modelling, Heterogeneous materials
\end{keyword}

\end{frontmatter}

\section{Introduction}

%
Reduced Order Modelling (ROM) has become an established way to accelerate numerical analyses by exploiting the information contained in previously obtained solutions (snapshots)---be it time-steps/load increments~\cite{ryckelynck_priori_2005} or solutions to other parametrizations~\cite{bolzon_effective_2011}---to construct an approximation space better suited to the investigated problem.
ROM is thus particularly appealing to problems such as optimization/parameter identification~\cite{bolzon_effective_2011}, system control~\citep{astrid_missing_2008}, or real-time simulations~\citep{barbic_real-time_2005,an_optimizing_2008,kim_physics-based_2011,harmon_subspace_2013,niroomandi_accounting_2012,radermacher_comparison_2013}, i.e. scenarios where the increased cost of an offline phase (collecting and processing snapshots) can be amortized in the subsequent calculations.
Many strategies in multi-scale modelling have a similar multi-query character; ROM thus allows for significant acceleration of $\textrm{FE}^2$-like computational homogenization approaches~\cite{yvonnet_reduced_2007,oliver_reduced_2017,kunc_finite_2019} and closely related applications in multi-scale topology optimisation~\cite{xia_reduced_2014,fritzen_topology_2016}.

\begin{figure*}[h!]
	\centering
	\setlength{\tabcolsep}{0pt}
	\begin{tabular}{>{\centering}m{0.30\textwidth} >{\centering}m{0.21\textwidth} >{\centering}m{0.48\textwidth} }
		\multicolumn{3}{c}{\includegraphics[width=0.975\textwidth]{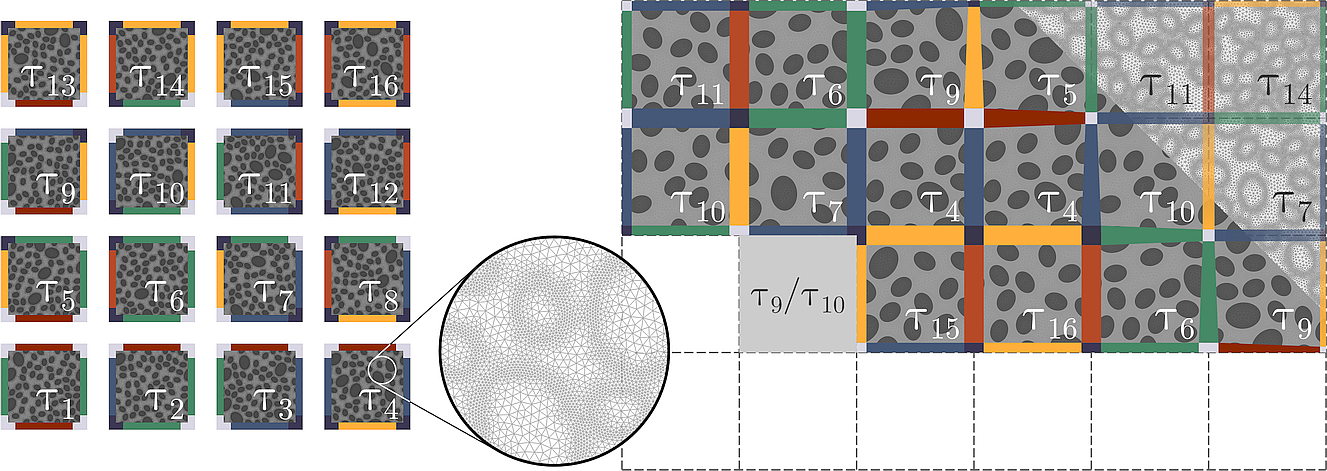}} \\
		(a) & (b) & (c)
	\end{tabular}	
	\caption{(a) A set of sixteen Wang tiles $\tau_{1},\dots,\tau_{16}$ with the compressed microstructural representation used in the work (including a detail of the finite element discretization in (b)) and (c) an illustration of a step in the tiling procedure (with partially assembled microstructural geometry and finite element discretization). In the next step of the assembly algorithm, either $\tau_9$ or $\tau_{10}$ will be placed at the position highlighted in grey.}
	\label{fig:concept_illustration}
\end{figure*}

\subsection{Local reduced models}
%
In general, the projection-based ROM works particularly well for problems with a relatively low-dimensional parametrization space (e.g. variation in geometrical or loading parameters, or prescribed macroscopic strain) and mild non-linearities~\cite{kerfriden_partitioned_2013,radermacher_model_2014}. 
To avoid a significant increase in dimensions of the reduced approximation spaces for more severe non-linearities, several strategies were proposed that approximate a solution manifold with multiple reduced approximation subspaces along with procedures allowing for interpolating/transition among them, e.g.~\cite{amsallem_nonlinear_2012,eftang_parameter_2012,peherstorfer_localized_2014}.
Besides these local approaches in terms of parametrization partitioning~\cite{eftang_parameter_2012} or solution characteristics~\cite{amsallem_nonlinear_2012,peherstorfer_localized_2014},
ROM can be utilized locally in problem's subdomains; for instance,
\citet{kerfriden_partitioned_2013} proposed using reduced modes only in regions without localized damage. 

In the FE$^{2}$ setting, \citet{oliver_reduced_2017} split a Representative Volume Element into a linear subdomain and a region of softening cohesive bands, extracting reduced modes for each subdomain separately.
\citet{radermacher_model_2014} introduced an adaptive sub-structuring ROM in forming process simulations, in which only the parts of the domain with mild non-linearity were reduced while highly non-linear regions remained fully resolved.
Similarly, \citet{niroomandi_real-time_2012} combined global ROM with the Generalized Finite Element Method (GFEM), introduced locally in a patch near simulated surgical cut, in order to capture localized solution characteristics efficiently. 
Using Partition of Unity (PU)~\cite{melenk_partition_1996}, underlying the previously mentioned GFEM approaches, \citet{ibanez_local_2019} recently proposed a local variant of Proper Generalized Decomposition.

Still, the parametrization space of problems with fully resolved, stochastic microstructural details is inherently too high-dimensional, see~\Sref{sec:standard_ROM} for an explicit example, and hence difficult to address within the standard ROM setting.
Even though \citet{kerfriden_partitioned_2013} adopted a ROM strategy for fracture modelling in domains with a stochastic microstructure, only the parameters of a damage model varied within the macroscopic domain. Consequently, the regions excluding the damage localization, which were accelerated with ROM, effectively behaved homogeneously in the linear regime.
Numerical strategies based solely on PU/GFEM thus seem to be better suited for simulations with stochastic microstructural details.

\subsection{Alternatives based on Partition of Unity}
Covering \emph{linear problems}, \citet{strouboulis_generalized_2003} introduced a dictionary of pre-computed local solutions to selected microstructural features, which were then used as enrichment functions within the Generalized Finite Element Method (GFEM).
\citet{fish_multiscale_2005} proposed a multiscale enrichment method, combining Partition of Unity and responses of a periodic microstructure representation to unit loading cases from the computational homogenization~\cite{geers_homogenization_2017}.
\citet{efendiev_generalized_2013} developed the Generalized Multiscale Finite Element Method which extracts local enrichments in problem's subdomains from a collection of pre-computed general responses using eigenvalue analyses. 
In a similar spirit but without pre-calculations, \citet{plews_bridging_2015} generated the microstructure-specific enrichments on-the-fly by subjecting subdomains with finer discretization to boundary values obtained from a coarse global solution, building on the local-global enrichment framework~\cite{duarte_analysis_2008}.
Outside the GFEM family of approaches, the recently proposed Coarse Mesh Condensation Multiscale method~\citep{le_coarse_2020}, subsequently extended to non-periodic microstructures and non-conforming coarse-scale discretization~\citep{le_full-field_2020}, is conceptually similar to the aforementioned as it combines a coarse scale approximation of a strain field with parallel calculations of localization fields within (potentially overlapping) subdomains. In contrast to older works of Zohdi and coworkers, e.g.~\cite{zohdi_domain_1999,zohdi_method_2001}, in which a domain-decomposition-like approach with a regularized approximation at subdomains' interfaces was adopted, the parametrization of the localization fields is linked to the coarse strain field via an $L_{2}$-norm projection in~\citep{le_coarse_2020,le_full-field_2020}.

\subsection{Our contribution}
%
%
In the series of our previous works~\cite{novak_compressing_2012,doskar_aperiodic_2014,doskar_level-set_2020}, we have introduced the framework of Wang tiles as a suitable extension of the (Statistically Equivalent) Periodic Unit Cell methodology for modelling microstructural geometry of random heterogeneous materials.
We have shown that replacing the unit cell based representation with a set of domains---Wang tiles---with predefined mutual compatibility enables an efficient generation of stochastic microstructural realizations that feature suppressed periodicity artefacts~\cite{novak_compressing_2012,doskar_aperiodic_2014,doskar_level-set_2020}
Even though the spurious periodicity is significantly reduced in the generated microstructural samples, they are still composed of only a handful of tiles. The Wang tile concept thus provides a finite-size discrete \emph{parametrization space} of all realizations.

Combining both GFEM and ROM approaches recalled above and benefiting from the discrete microstructure parametrization, we develop a reduced order modelling scheme for problems with fully resolved microstructural details generated by the tile-based approach. 
We first recall the essentials of Wang tile concept and its use in modelling heterogeneous materials in~\Sref{sec:wang_tiles_introduction}. Next, inspired by computational homogenization approaches as in~\cite{fish_multiscale_2005}, we propose a method for extracting the characteristic responses of the compressed microstructural representation to parametrized macroscopic loading in~\Sref{sec:modes_extraction}. These general, tile-wise defined responses are constructed such that they can be assembled in the same way a microstructural sample is synthesized from the individual Wang tiles, and they serve as an approximation to the fluctuation part of a macroscopic solution.
\Sref{sec:macroscopic_scheme} then introduces the macroscopic numerical scheme that combines the general pre-computed characteristic responses using the Generalized Finite Element ansatz to generate the microstructure-informed reduced modes, specific for a given macroscopic problems.
We illustrate the proposed methodology with two-dimensional, scalar, elliptic examples in~\Sref{sec:numerical_examples}, and discuss the obtained results in~\Sref{sec:conclusions}.
Finally, we summarize the proposed scheme and outline its possible extensions in~\Sref{sec:summary}.

\section{Wang tiles as microstructural ROM}
\label{sec:wang_tiles_introduction}

%
The concept of Wang tiles was originally proposed as an equivalent problem in predicate calculus~\citep{wang_proving_1961}: instead of proving a logical statement of a certain class directly, it was converted to a question whether a set of square tiles with codes attributed to their edges can tile an infinite plane such that the adjacent tiles have the same code on the corresponding edges. 
Later, this abstract concept was applied in the Computer Graphics community as a convenient formalism for encoding continuity constraints when assembling pre-generated modules into larger blocks---tilings. Adopting the tile concept has enabled fast synthesis of point patterns with desired blue noise spectrum~\cite{hiller_tiled_2001,kopf_recursive_2006} and naturally looking textures~\cite{cohen_wang_2003,sibley_wang_2004,zhang_efficient_2008}.
The latter applications inspired utilisation of the tile concept in modelling materials with random heterogeneous microstructures~\cite{novak_compressing_2012}.

\begin{figure*}[h!]
	\setlength{\tabcolsep}{2pt}
	\begin{tabular}{ccc}
		\includegraphics[width=0.32\textwidth]{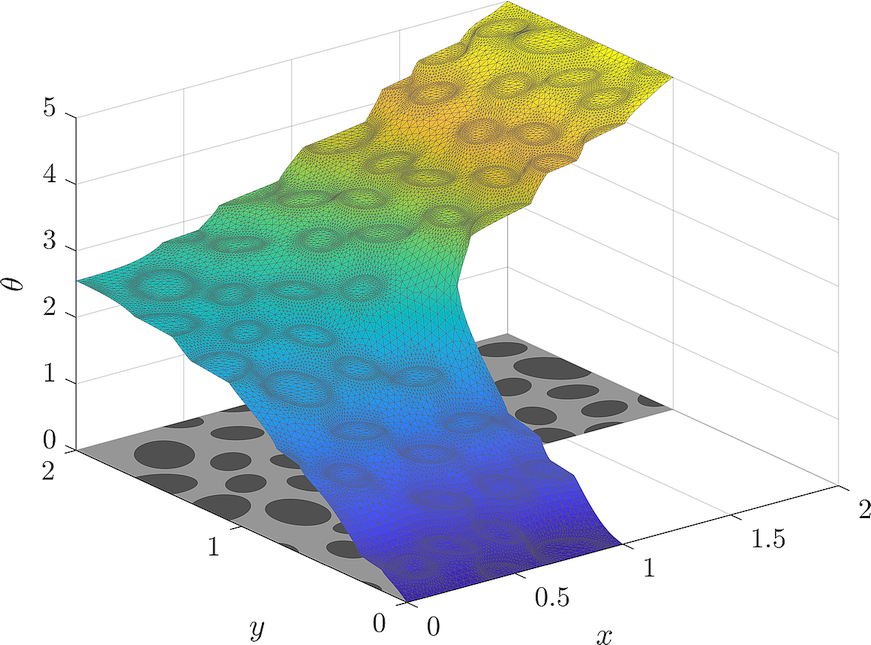} & \includegraphics[width=0.32\textwidth]{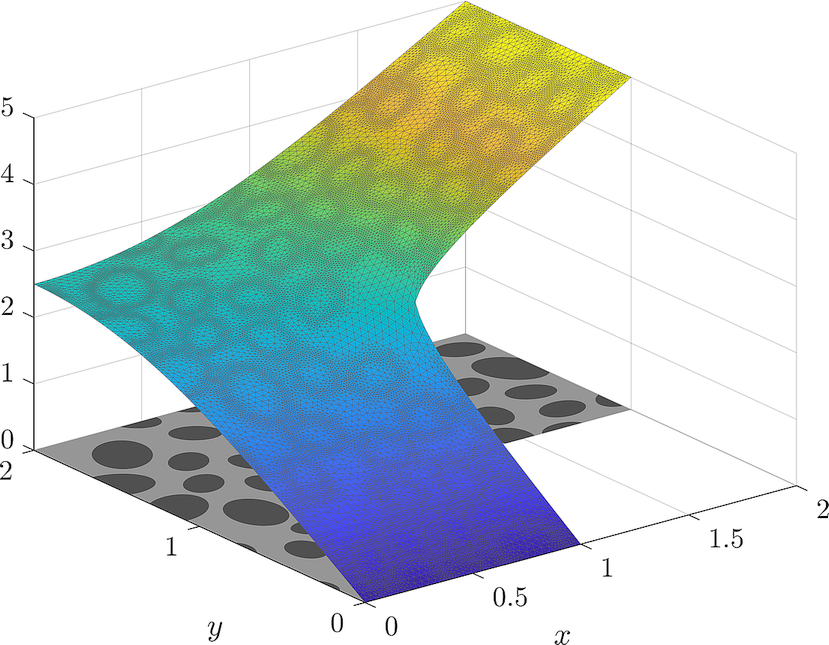} &\includegraphics[width=0.32\textwidth]{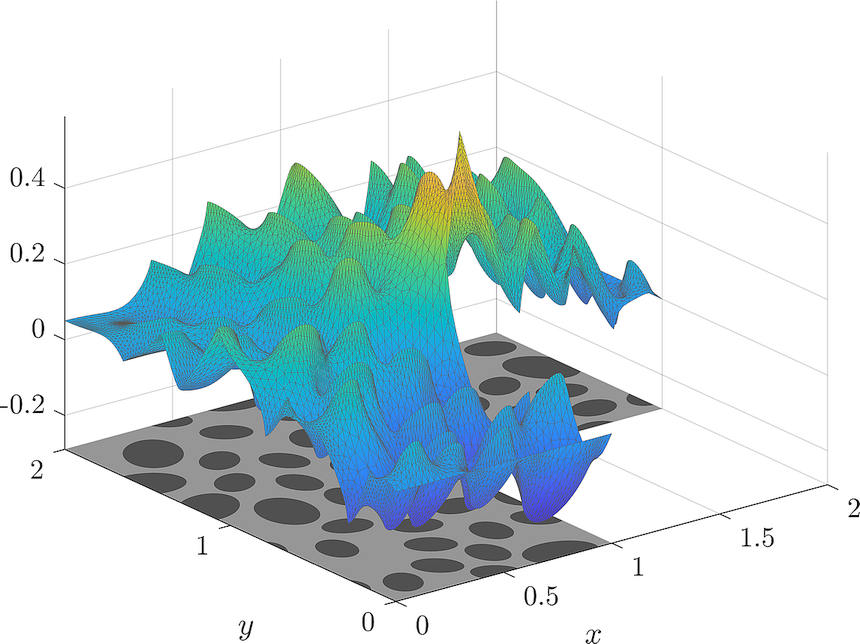} \\
		(a) & (b) & (c)
	\end{tabular}
	\caption{Decomposition of (a) a Direct Numerical Solution resolving all microstructural details into (b) a homogeneous part and (c) fluctuations caused by the presence of heterogeneities.}
	\label{fig:illustrate_fluctuation}
\end{figure*}

%
Currently, characterization of a microstructural geometry in a compressed form is mostly based on a Periodic Unit Cell (PUC) representation (under different names such as Repeating Unit Cell~\cite{yang_new_2018}, Statistically Similar Representative Volume Element~\cite{balzani_construction_2014}, Statistically Optimal Representative Unit Cell~\cite{lee_three-dimensional_2009}, or Statistically Equivalent Periodic Unit Cell~\cite{zeman_random_2007}).
Although such compression is lossless only for a narrow range of materials with a periodic arrangement of constituents, the modelling error caused by the periodicity and limited variability is negligible in standard concurrent multiscale schemes, such as FE$^2$~\cite{feyel_fe2_2000}.
On the other hand, the PUC-based compression is unsuitable for generating microstructural geometry for macroscale analyses with a fully resolved microstructure for two reasons: (i) it produces only a single, hence deterministic, sample and (ii) the produced microstructural sample exhibits artificial periodicity. 
Wang tile concept addresses both drawbacks, and allows to avoid computationally intensive optimization algorithms traditionally used to generate an ensemble of stochastic microstructural samples~\cite{zeman_random_2007}.

%
In the Wang tile (cube) concept, a material microstructure is represented with a set of square (or cubical) domains with codes attributed to their edges (or faces). Interior of each tile is designed such that (i) the contained microstructural geometry remains continuous across the parts of tile boundary with the same code and (ii) when assembled following the compatibility constraints posed by the codes, the resulting tile assembly resembles the original microstructure in terms of microstructural features, assessed typically via spatial statistics; see the illustration in~\Fref{fig:concept_illustration}. 

So far, three design strategies have been proposed: an optimization approach to minimize a discrepancy between spatial statistics of the reference specimen and generated assemblies~\cite{novak_compressing_2012}, a sample-based strategy~\cite{doskar_aperiodic_2014}, and a level-set based framework for particulate and foam-like microstructures~\cite{doskar_level-set_2020}. However, most of the methods developed for generating PUC can in principle be modified to adopt the generalized periodicity constraints of Wang tiles.

Unlike mathematicians that focus mainly on tile sets for strictly aperiodic tiling, see the overview in~\cite[Chapter 11]{grunbaum_tilings_2016}, we prefer stochastic tile sets introduced by Cohen et al.~\cite{cohen_wang_2003} because these sets perform better in terms of suppressing periodic artefacts~\cite{doskar_aperiodic_2014,doskar_level-set_2020} and are less constrained in their construction.
Using the assembly algorithm described next, a sufficient condition for a tile set is that it contains at least one tile for each admissible code combination on the left and top tile edges. Otherwise, the number of tiles and code distribution within the set can be chosen arbitrarily, e.g., to control the distribution of individual tile types in the assembled tilings.
Individual microstructural samples are generated with a stochastic assembly algorithm that sequentially fills a grid of a predefined size with tile instances from the set. For each position in the grid, the algorithm identifies potential candidate tiles from the set based on the codes of previously placed neighbouring tiles, randomly selects one tile from the candidates, places it, and proceeds to the next grid position. A step of the assembly algorithm is shown in~\Fref{fig:concept_illustration}.

From the perspective of Liu and Shapiro~\cite{liu_random_2015}, who define a model of a material microstructure as a process capable of generating microstructural samples with similar spatial statistics, Wang tile concept can be seen as a reduced model with spatially local modes (with PUC being a trivial instance with one mode only).
Characteristic microstructural features are compressed into the set of tiles in the offline phase, while geometries of individual microstructural samples are generated almost instantly when required.
The tile-based representation thus can be advantageously used in. e.g., analyses of the Representative Volume Element (RVE) size~\cite{doskar_jigsaw_2016,doskar_wang_2018}.
The concept can also serve as a microstructural generator for Monte-Carlo-based simulations investigating the influence of microstructural variability on a macroscopic response.

\section{Extracting characteristic fluctuation fields}
\label{sec:modes_extraction}


%
The idea of approximating fluctuation fields in microstructural samples generated by means of Wang tiles was first outlined in~\citet{novak_microstructural_2013}; see~\Fref{fig:illustrate_fluctuation} for an illustration. Aiming at stress enrichments for Hybrid-Trefftz finite element formulation such as~\cite{novak_micromechanics-enhanced_2012}, Novák and coworkers proposed extracting tile-wise defined fields from a response of a selected tiling to a prescribed macroscopic strain under periodic boundary conditions. Continuity of traction forces across the corresponding tile edges was incorporated in the objective function during the optimisation-based tile design, resulting in a trade-off between randomness in the system and traction jumps~\cite{novak_microstructural_2013}.

Here, we present a method for extracting the characteristic response of a compressed system in primal variables. Unlike~\cite{novak_microstructural_2013}, the tile-wise defined fields are continuous across the corresponding edges \emph{by construction} and the method is non-intrusively applicable to allmexisting microstructural compressions.
Our only assumption is that the finite element discretization of individual tile domains $\domain^{\tiledom{}}$ is geometrically compatible across the edges with the same code.

%
We illustrate the method with a scalar, elliptic problem represented by heat conduction. 
Combining the Fourier law
\begin{equation}
	\flux\atx = - \tenss{K}\atx \grad \temperature\atx \,,\quad\forall\x\in\domain\,,
\end{equation}
which links temperature $\temperature$ and heat flux $\flux$ fields in a given domain $\domain$ via a conductivity tensor $\tenss{K}$, with the conservation of the heat flux (neglecting any heat sources and sinks)
\begin{equation}
 \divergence \flux\atx = 0	
\end{equation}
yields the governing equation
\begin{equation}
	- \divergence \left( \tenss{K}\atx \grad \temperature\atx \right) = 0\,.
	\label{eq:governing}
\end{equation}

Similarly to the ansatz traditionally used in computational homogenization, we assume that a solution field for each tile $\tiledom{}$ in a tile set $\setdom$ can be decomposed, using a Taylor expansion, into a macroscopic part controlled by a prescribed macroscopic (potentially higher-order) gradients and the fluctuation part caused by the presence of heterogeneities in the microstructure.
Restricting ourselves to the second-order Taylor polynomial, we assume the temperature field $\temperature\atx$ in the form
\begin{equation}	
	\temp\atx = \tGradF \scontr \x + \half \x \scontr \tGradS \scontr \x + \tFluct\atx\,,\,\, \forall \x \in \domain^{\tiledom{}},\,\, \forall \tiledom{} \in \setdom,
	\label{eq:expansion}
\end{equation}
where $\tGrad{i}$, $i\in\left\{1,2\right\}$, is the prescribed $i$th-order macroscopic gradient playing the role of generalized loading and $\tFluct\atx$ is the fluctuation complement.

Following the standard finite element procedures for solving the weak form of~\Eref{eq:governing}, we construct a stiffness matrix for each tile independently, and we reorder and split the unknowns $\sevek{u}$ into the interior, $\sevek{u}_{i}$, and exterior, $\sevek{u}_{e}$, part, resulting in a set of linear equations for each tile $\tiledom{}$ in the form:
\begin{equation}
	\begin{bmatrix}
		\semtrx{K}_{ee}^{\tiledomg} & \semtrx{K}_{ei}^{\tiledomg} \\ \semtrx{K}_{ie}^{\tiledomg} & \semtrx{K}_{ii}^{\tiledomg}
	\end{bmatrix}
	\begin{bmatrix}
		\sevek{u}_{e}^{\tiledomg} \\
		\sevek{u}_{i}^{\tiledomg} 
	\end{bmatrix}
	=	
	\begin{bmatrix}
		\sevek{f}_{e}^{\tiledomg} \left( \tGradF, \tGradS \right)\\
		\sevek{f}_{i}^{\tiledomg} \left( \tGradF, \tGradS \right)
	\end{bmatrix}
	\,.
	\label{eq:tile_problem}
\end{equation}
Note that the force term is induced by the prescribed macroscopic gradients and the degrees of freedom (DOFs) correspond to the fluctuation part of the ansatz in~\Eref{eq:expansion}.
The interior DOFs are then condensed out for each tile, leaving us with load vector $\hat{\sevek{f}}^{\tiledomg}$ and effective stiffness matrix $\hat{\semtrx{K}}^{\tiledomg}$ as the Schur complement of $\semtrx{K}_{ii}^{\tiledomg}$ from the original problem~(\ref{eq:tile_problem}):
\begin{align}
	\hat{\semtrx{K}}^{\tiledomg} &= \semtrx{K}_{ee}^{\tiledomg} - \semtrx{K}_{ei}^{\tiledomg} \, {\semtrx{K}_{ii}^{\tiledomg}}^{-1} \, \semtrx{K}_{ie}^{\tiledomg} \,,\\
	\hat{\sevek{f}}^{\tiledomg} &= \,\sevek{f}_{e}^{\tiledomg} - \semtrx{K}_{ei}^{\tiledomg} \, {\semtrx{K}_{ii}^{\tiledomg}}^{-1} \, \sevek{f}_{i}^{\tiledomg} \,,
\end{align}
where the explicit dependence on the macroscopic gradients has been omitted for the sake of brevity.

To enforce continuity of the solution across individual tiles, the remaining boundary DOFs are renumbered and assembled into $\sevek{u}^{s}$ such that the corresponding DOFs at edges with the same code share the same number throughout the tile set.
While uniquely enumerating DOFs that belong exclusively to a tile edge is straightforward, enumerating vertex DOFs requires an analysis of the tile set described in~\cite[Section 2.3]{doskar_level-set_2020} in order to identify how many unique DOFs are needed for the vertices.
Depending on the code distribution within the set, it may happen that certain vertices will never coincide during the tile assembly, thus constituting separate vertex group with distinct DOFs.
This typically happens for tile sets derived from the vertex-defined Wang tiles~\cite{lagae_alternative_2006}.
In the three-dimensional setting, where individual Wang tiles are defined by codes attributed to domain faces, a similar analysis must be performed also for edge DOFs; see \cite[Section 2.3]{doskar_level-set_2020} for details.

Finally, treating each tile as a macro-element, the effective tile stiffness matrices and loading vectors are assembled according to the DOF numbering,
\begin{equation}
	\semtrx{K}^{\setdom} = \assembly_{\tiledomg\in\setdom} \hat{\semtrx{K}}^{\tiledomg} 
	\quad\text{and}\quad
	\sevek{f}^{\setdom} = \assembly_{\tiledomg\in\setdom} \hat{\sevek{f}}^{\tiledomg} \,,
\end{equation}
forming a system that encodes a collective response of all tiles in the set to a prescribed loading,
\begin{equation}
	\semtrx{K}^{\setdom} \sevek{u}^{\textrm{s}} = \sevek{f}^{\setdom} \,.
	\label{eq:set_system}
\end{equation}
Without loss of generality, we assume all tile domains are centred, making them virtually stacked one atop each other. However, their interaction is facilitated only through the shared DOFs.

\begin{figure*}[ht!]
	\centering
	\includegraphics[width=0.85\textwidth]{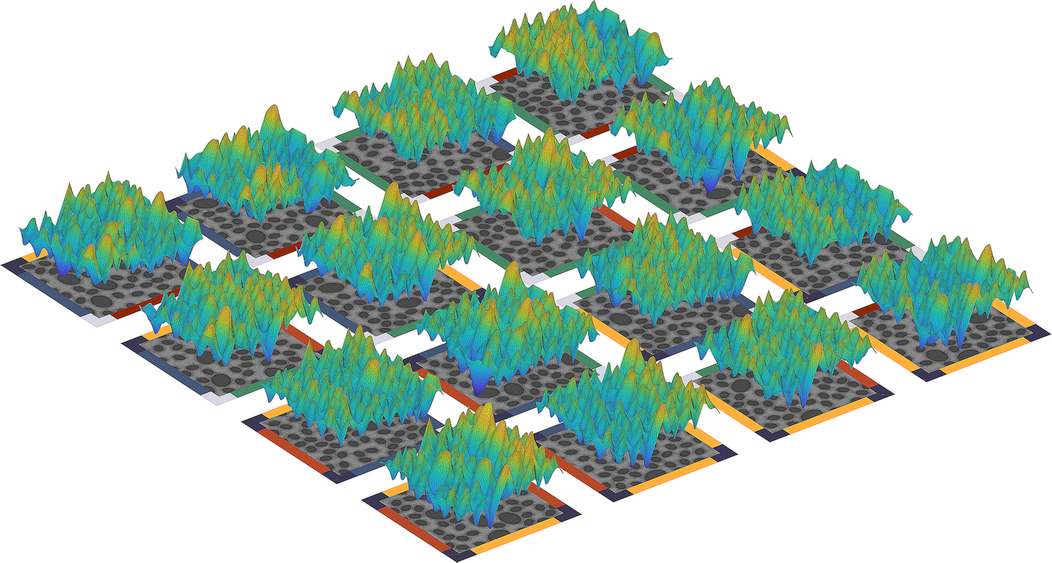}
	\caption{An example of a tile-wise defined fluctuation field obtained for $\tGradF = \protect\begin{bmatrix}1.0&0.0\protect\end{bmatrix}\trn$ and considering boundary conditions (\ref{eq:1st_PBC}).}
	\label{fig:field_example}
\end{figure*}

\subsection{Constraints}
\label{sec:modes_extraction_constraints}

The assembled set problem, \Eref{eq:set_system}, must be equipped with additional constraints. First, the problem is ill-posed because it contains a zero-energy mode. This can be fixed either by prescribing a value for one DOF or by imposing the average value constraint
\begin{equation}
	\int_{\domain^{\setdom}} \tFluct \de{\x} = 0\,,
	\label{eq:zero_average_constraint}
\end{equation}
where $\domain^{\setdom} = \bigcup_{\tiledomg{}\in\setdom} \domain^{\tiledomg{}}$.

%
Second, additional constraints are needed to prevent the fluctuation field from compensating for the prescribed loading. This corresponds to prescribing suitable boundary conditions at the microscale problem in the computational homogenization, e.g.~\cite{geers_homogenization_2017}.
 
In the standard first-order computational homogenization, the fluctuation field $\tFluct$ is required to satisfy
\begin{equation}
	\int_{\domain} \grad \tFluct \de{\x} = \int_{\boundary} \tFluct \tens{n} \de{s} = \tens{0} \,,
	\label{eq:vanishing_fluct_grad}
\end{equation}
where $\tens{n}$ denotes the outer normal to the RVE domain boundary $\boundary$.
This is typically ensured either with the uniform Dirichlet, periodic, or uniform Neumann type of boundary conditions. Albeit the particular selection is a modelling choice, periodic boundary conditions are the most frequent due to their fastest convergence of effective properties with respect to the size of an RVE, e.g.~\cite{kanit_determination_2003}.

Loosely inspired, we implement the following three options, in which we
\begin{enumerate}[i)]
	\item set all fluctuation DOFs to zero at the tile boundaries, 
	\begin{equation}
		\tFluct\at{\partial\domain^{\tiledomg}} = 0\,, \quad \forall \tiledomg \in \setdom \,;
		\label{eq:1st_DBC}
	\end{equation}
	
	\item enforce vanishing gradient of the fluctuation field tile-wise,
	\begin{equation}
		\int_{\partial\domain^{\tiledomg}} \tFluct\tens{n} \de{s} = \tens{0}\,, \quad \forall \tiledomg \in \setdom \,;
		\label{eq:1st_PBC}
	\end{equation}
	
	\item require zero gradient of $\tFluct$ within the tile set on average,
	\begin{equation}
		\int_{\partial\domain^{\setdom}} \tFluct\tens{n} \de{s} = \tens{0}\,,
		\label{eq:1st_NBC}
	\end{equation}
	where $\partial\domain^{\setdom} = \bigcup_{\tiledomg{}\in\setdom} \partial\domain^{\tiledomg}$, similarly to~\Eref{eq:zero_average_constraint}.
\end{enumerate}

%
Treating second-order fluctuations is more involved.
Since we assume the classical first-order constitutive and conservation equations and we want to pose the extraction of characteristic modes as a boundary value problem, we cannot straightforwardly enforce a direct analogue of \Eref{eq:vanishing_fluct_grad} for the second-order gradient. 
Addressing this limitation in the finite strain setting, Kouznetsova and coworkers~\cite{kouznetsova_multi-scale_2002,kouznetsova_multi-scale_2004} derived a constraint on the fluctuation part of the solution that ensures the equality between the prescribed second-order gradient and its average over a Periodic Unit Cell, see~\cite[Appendix A]{kouznetsova_multi-scale_2002}.
Adapted to the illustrative problem of heat conduction, Kouznetsova et al.'s condition reads
\begin{equation}
	\int_{\boundary} \tFluct \left( \x \otimes \tens{n} + \tens{n} \otimes \x \right) \de{s} = \tenss{0} \,,
	\label{eq:vanishing_second_fluct_grad}
\end{equation}
where $\otimes$ denotes the vector outer product defined in the index notation as $(\tens{n} \otimes \x)_{ij} = n_i x_j	$.

Similarly to the previous paragraph, constraint~(\ref{eq:vanishing_second_fluct_grad}) can be ensured
\begin{enumerate}[i)]
	\item point-wise (leading to the same constraints as in \Eref{eq:1st_DBC}),
	\item tile-wise
	\begin{equation}
	\int_{\partial\domain^{\tiledomg}}  \tFluct \left( \x \otimes \tens{n} + \tens{n} \otimes \x \right) \de{s} = \tenss{0}\,, \quad \forall \tiledomg \in \setdom \,,
	\label{eq:2nd_PBC}
	\end{equation}
	\item set-wise
	\begin{equation}
	\int_{\partial\domain^{\setdom}}  \tFluct \left( \x \otimes \tens{n} + \tens{n} \otimes \x \right) \de{s} = \tenss{0}\,.
	\label{eq:2nd_NBC}
	\end{equation}
\end{enumerate}
Note that, unlike the finite strain problem in~\cite[Appendix A]{kouznetsova_multi-scale_2002}, the second order gradient for the scalar problem is by definition symmetric. Hence, conditions (\ref{eq:2nd_PBC}) and (\ref{eq:2nd_NBC}) constitute only three constraints for a two-dimensional scalar problem.

\begin{figure*}[ht!]
	\setlength{\tabcolsep}{2pt}
	\begin{tabular}{ccc}
		\includegraphics[width=0.32\textwidth]{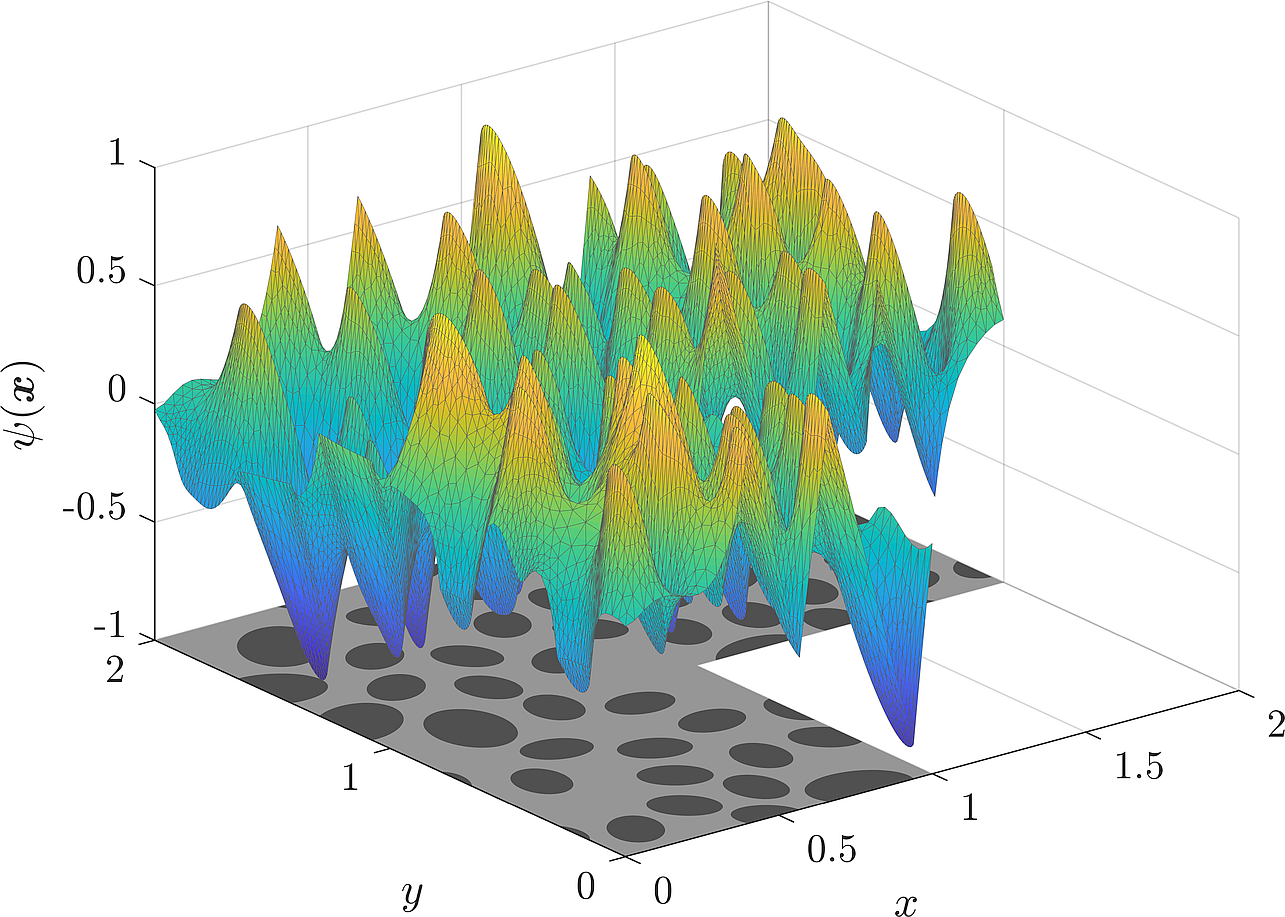} & \includegraphics[width=0.32\textwidth]{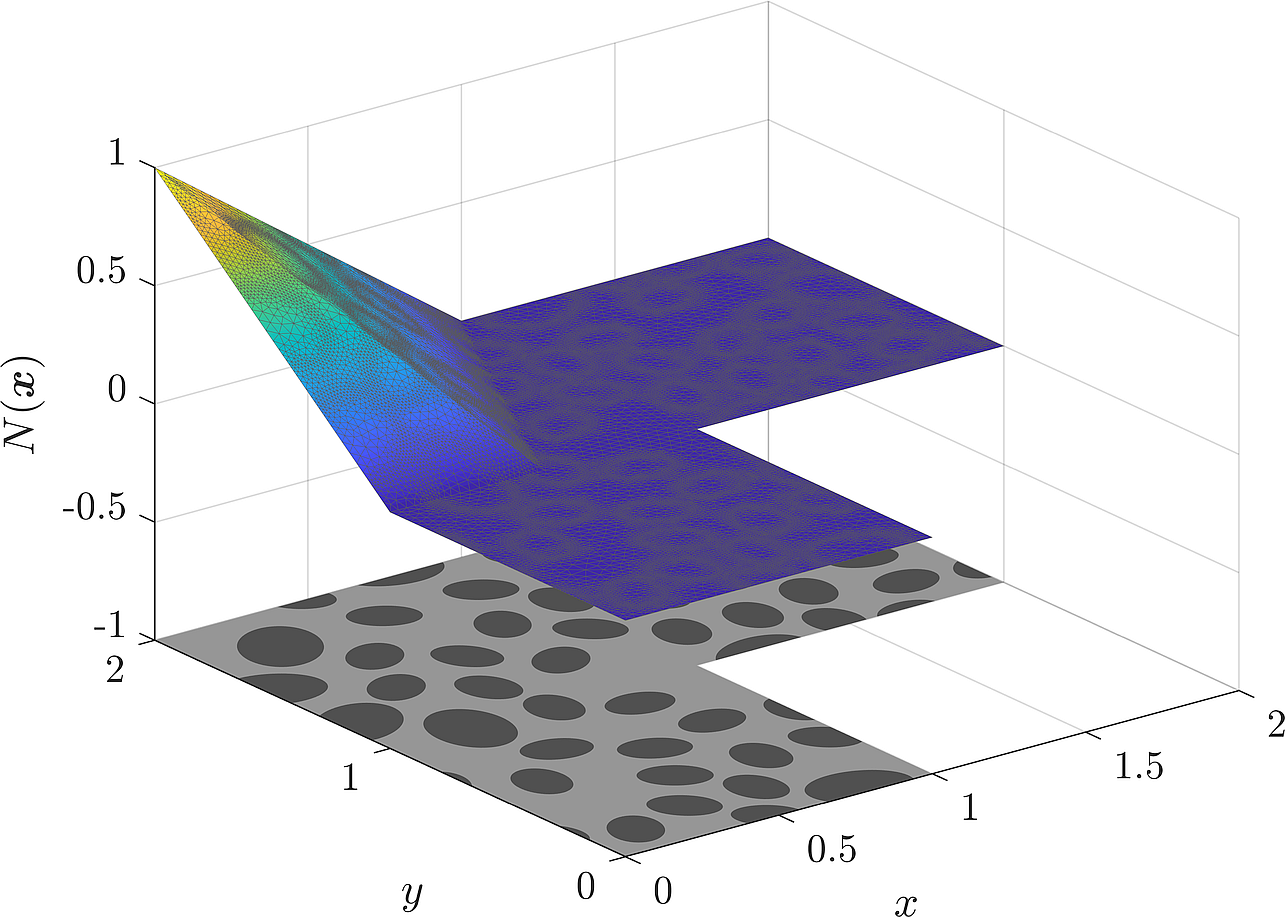} &\includegraphics[width=0.32\textwidth]{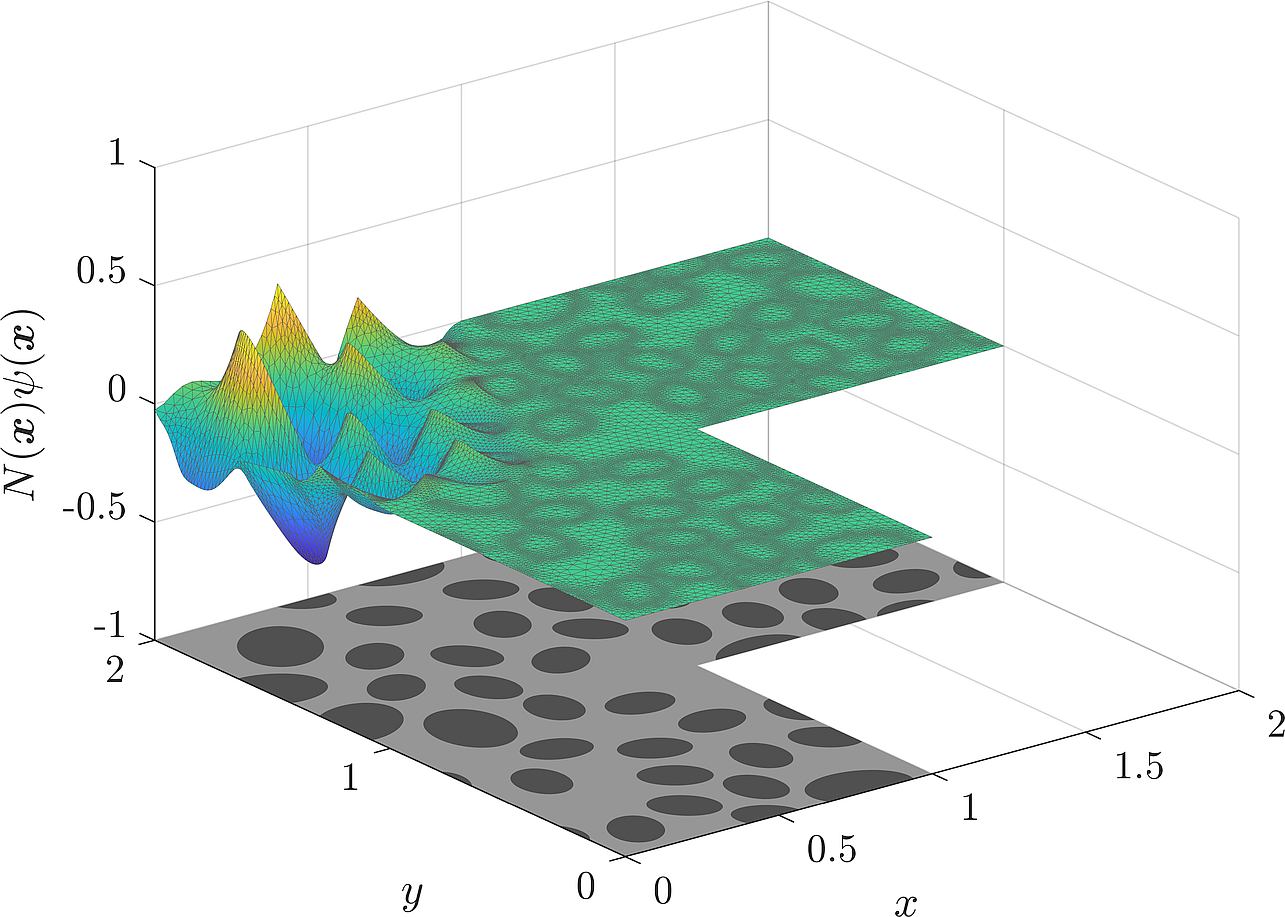} \\
		(a) & (b) & (c)
	\end{tabular}
	\caption{Illustration of (a) a global approximation $\psi\atx$ assembled from the tile-wise defined fluctuation field, (b) a standard macroscopic shape function $N\atx$ used as a Partition of Unity function in the Generalized Finite Element Method, and (c) the final microstructure-informed reduced mode $N\atx\psi\atx$.}
	\label{fig:GFEM_combination}
\end{figure*}

%
All constraints introduced above are linear in the fluctuation field $\tFluct$. Therefore, within the finite-element implementation, they can be recast to the form
\begin{equation}
	\semtrx{C} \sevek{u} = \sevek{0} \,, \quad \text{with} \quad \semtrx{C} = \begin{bmatrix} \semtrx{C}_{\text{0}} \\ \semtrx{C}_{\text{I}} \\ \semtrx{C}_{\text{II}} \end{bmatrix} ,
\end{equation}
where $\semtrx{C}_{\text{0}}$, $\semtrx{C}_{\text{I}}$, and $\semtrx{C}_{\text{II}}$ are related to the elimination of the zero-energy mode, and enforcing first- and second-order gradients, respectively.

%
However, depending on the particular constraint choice and the distribution of edge codes with the tile set, matrix $\semtrx{C}$ can be rank-deficient. This deficiency arises in several situations.
For instance, a tile with periodic codes (i.e. the same codes on the opposite edges) automatically satisfies~\Eref{eq:1st_PBC}.
In a similar manner, a tile set with an equal occurrence of individual codes on opposite edges yields trivial $\semtrx{C}_{\text{I}}$ for~\Eref{eq:1st_NBC}.

In principle, it is possible to list all such situations and handle them during the assembly of matrix $\semtrx{C}$. Yet, in order to avoid an involved analysis of codes in a tile set and complicated assembly, we perform the singular value decomposition of the rank-deficient matrix $\semtrx{C}$ and replace it with $\hat{\semtrx{C}}$ containing the right-singular vectors related to non-zero singular values as its rows.

%
Ultimately, the modified constraints represented by $\hat{\semtrx{C}}$ are enforced via Lagrange multipliers $\lambda$, forming the final saddle-point problem
\begin{equation}	
	\begin{bmatrix} \semtrx{K}^{\setdom} & \hat{\semtrx{C}}\trn \\ \hat{\semtrx{C}} & \semtrx{0} \end{bmatrix}
	\begin{bmatrix} \sevek{u}^{\textrm{s}} \\ \sevek{\lambda} \end{bmatrix}
	= 
	\begin{bmatrix} \sevek{f}^{\setdom} \\ \sevek{0} \end{bmatrix}\,.
	\label{eq:saddle-point_system}
\end{equation}
Once solved for the exterior DOFs $\sevek{u}^{\textrm{s}}$ common to all tiles, the remaining interior DOFs follow from back-substitution\footnote{In the case of zero fluctuations prescribed at tile boundaries, i.e.~\Eref{eq:1st_DBC}, the above described treatment is only formal. In practice, solving~\Eref{eq:saddle-point_system} is skipped, and only the interior DOFs are sought for because no information is effectively communicated across individual tiles.} in~\Eref{eq:tile_problem},
\begin{equation}
	\sevek{u}_{i}^{\tiledomg} = {\semtrx{K}_{ii}^{\tiledomg}}^{-1} \left( \sevek{f}_{i}^{\tiledomg} - \semtrx{K}_{ie}^{\tiledomg} \sevek{u}_{e}^{\tiledomg} \right) \,.
\end{equation}

\subsection{Field options}
\label{sec:mode_families}

%
In the illustrative linear two-dimensional problem, five basic macroscopic loading cases suffice, consecutively prescribing a unit value to the individual components of the first-, $\tGradF{}$, and second-order, $\tGradS{}$, macroscopic gradient, respectively.
Each load case is further complemented with one of the constraint options mentioned above.
Unlike in computational homogenization, where a particular type of boundary conditions is assumed a priori, we combine the characteristic responses of the compressed system extracted for different constraint options to enrich the basis of the fluctuation approximation space.

While this approach allows for a range of combinations regarding which gradients are prescribed and restricted, we consider only three choices for the rest of the paper, labelled as $1^{\text{st}}$, $1^{\text{st}}\vee2^{\text{nd}}$, and $1^{\text{st}}\wedge2^{\text{nd}}$, respectively.
The first choice denotes the fluctuation fields extracted by prescribing and constraining only the first-order macroscopic gradient, i.e. $\tGradS = \tenss{0}$ and only the constraints (\ref{eq:1st_DBC})--(\ref{eq:1st_NBC}) are considered. 
The second choice stands for the case where the first and second-order macroscopic gradients were prescribed independently, i.e. extending the set of fields from the previous choice with additional ones obtained similarly by setting $\tGradF = \tens{0}$ and prescribing only the second-order gradient along with constraints (\ref{eq:1st_DBC}), (\ref{eq:2nd_PBC}), or (\ref{eq:2nd_NBC}).
Finally, the last choice corresponds to the setup in which both gradients are prescribed simultaneously and all constraints are successively imposed.

The first option thus results in six fluctuation fields (arising from two unit loading cases and three constraint types), while the latter two yield 15 fluctuation fields (two unit loading cases for the first- and three for the symmetric second-order macroscopic gradient combined again with three constraint types). 

%

\section{Macroscopic numerical scheme}
\label{sec:macroscopic_scheme}

\begin{figure*}[h!]
	\centering
	\begin{tabular}{cc}
		\includegraphics[width=0.45\textwidth]{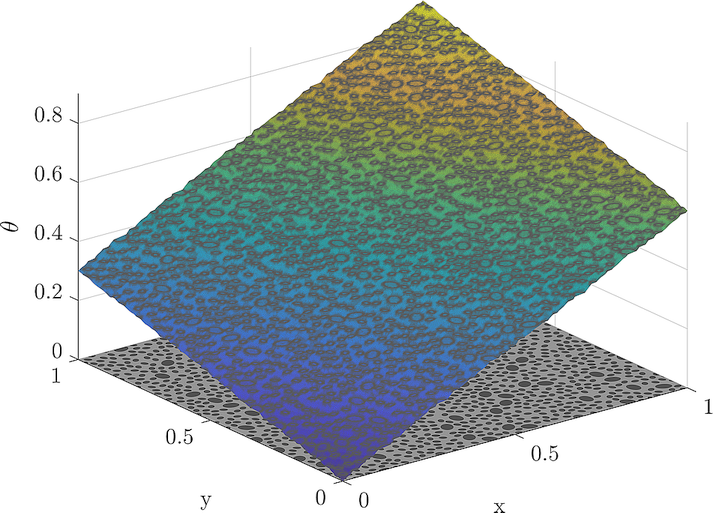} & \includegraphics[width=0.45\textwidth]{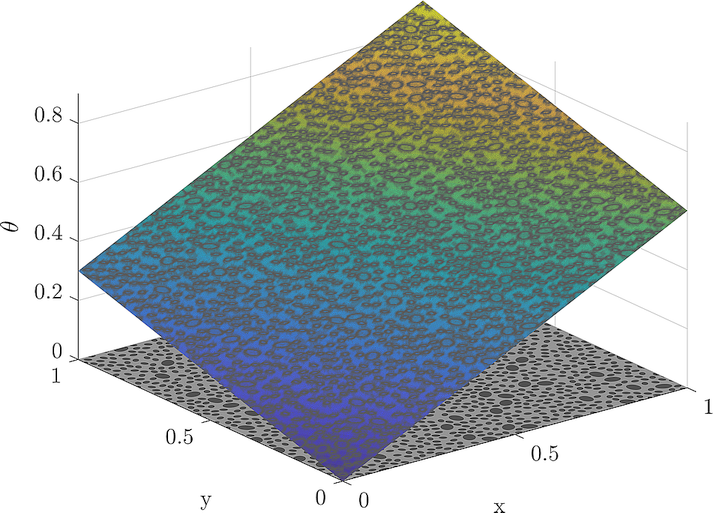} \\
		(a) & (b)
	\end{tabular}
	\caption{The resulting temperature field $\temp$ from a Direct Numerical Simulation (DNS) of a square domain with a microstructure assembled from $5\times5$ tiles and subjected to a macroscopic temperature gradient $\protect\begin{bmatrix}0.6&0.3\protect\end{bmatrix}\trn$ under (a) periodic boundary conditions and (b) Dirichlet boundary conditions.}
	\label{fig:simple_domain_illustration}
\end{figure*}

From the procedure outlined in the previous section, each tile carries a set of tile-wise defined fields, which can be assembled using the same procedure as in generating microstructural geometry in order to define a set of continuous approximation basis functions $\psi_i$ in the macroscopic domain $\domain$; see~\Fref{fig:GFEM_combination}
Note that the tile-wise defined fields are universal in the sense that they are not extracted with respect to a particular macroscopic geometry, arrangement of tiles in $\domain$, or loading; each $\psi_i$ approximates the microstructural response to a constant or linear macroscopic gradient.
However, the general macroscopic problem (with arbitrary geometry and loading) usually does not yield a uniform macroscopic gradient in the domain. Thus, the macroscopic numerical scheme must be able to locally interpolate between individual $\psi_i$ basis functions.

To facilitate such an interpolation, we resort to the Generalized Finite Element Method (GFEM) strategy~\cite{strouboulis_generalized_2001,belytschko_review_2009,fries_extended/generalized_2010}. We assume two levels of finite element discretization for the macroscopic problem.
The first level comprises the fine discretization that comes directly from the tile-wise discretization already used to pre-compute the tile characteristic fluctuation fields. Recall that the individual tile discretizations must be geometrically and topologically compatible for the strategy outlined in~\Sref{sec:modes_extraction} to work. Hence, these discretizations can be assembled along with the microstructural geometry, yielding $n^f$ shape functions $N^f\atx$ defining a finite element approximation space $\mathcal{U}^{f}=\text{span}\{N_i^{f}\}$, in which $\psi_i$'s are defined, i.e. $\text{span}\left\{\psi_i\right\} \subset \mathcal{U}^{f}$.

The second level contains a coarse discretization of the macroscopic domain $\domain$ that does not resolve microstructural features. With $n^{c}$ shape functions $N_i^{c}\atx$, it serves two purposes: (i) the related approximation space $\mathcal{U}^{c}$, $\mathcal{U}^{c} = \text{span}\left\{N_i^{c}\right\}$, captures the homogeneous part of the solution (recall that $\psi_i$ approximate only fluctuations caused by material heterogeneity), and (ii) the shape functions form a Partition of Unity basis for interpolating between individual $\psi_i$'s.

Following the GFEM ansatz, we seek a solution to a macroscopic problem in the form
\begin{align}
	\begin{split}
	\temp\atx 	&= \sum_{i=1}^{n^{c}} N_i^{c}\atx a_i^0  + \sum_{i=1}^{n^{c}} \sum_{j=1}^{n^{r}} N_i^{c}\atx \psi_j\atx a_i^j \\
				&= \sum_{i=1}^{n^{c}} \sum_{j=0}^{n^{r}} N_i^{c}\atx \psi_j\atx a_i^j \,, \quad \x\in\domain\,,
	\end{split}
	\label{eq:GFEM_ansatz}
\end{align}
where $n^r$ denotes the number of provided fluctuation fields and $\psi_0$ is artificially defined as a constant unit function.

In order to avoid difficulties common to GFEM, namely the questions of imposing essential boundary conditions and numerical integration, we adopt the perspective of Reduced Order Modelling. 
We benefit from the availability of the fine discretization of the macroscopic domain; plugging the fine discretization in the form
\begin{equation}
	\theta\atx = \sum_{i=1}^{n^{f}} N_i^{f}\atx u_i \,, \quad\x\in\domain\,,
\end{equation}
into the weak form of~\Eref{eq:governing} yields the standard, fine-scale system of linear equations
\begin{equation}
	\semtrx{K}^{f} \sevek{u} = \sevek{f}^{f} \,.
	\label{eq:full_system}
\end{equation}
However, instead of solving~\Eref{eq:full_system}, we approximate
\begin{equation}
	\sevek{u} \approx \semtrx{\Phi} \, \sevek{a}\,,
\end{equation}
where the matrix $\semtrx{\Phi} \in \setR^{n^{f}\times n^{c}(n^{r}+1)}$ collects as its columns the individual products $ N_i^{c}\atx \psi_j\atx$ from~\Eref{eq:GFEM_ansatz} projected onto $\mathcal{U}^{f}$, and $\sevek{a}$ contains the related $a_i^j$ DOFs.
The final reduced system to solve thus follows from the Galerkin projection,
\begin{equation}
	\underbrace{\semtrx{\Phi}\trn \semtrx{K}^{f} \semtrx{\Phi}}_{\semtrx{K}^{r}} \,\, \sevek{a} = \underbrace{\semtrx{\Phi}\trn \sevek{f}^{f}}_{\sevek{f}^{r}} \,,
	\label{eq:reduced_system}
\end{equation}
where the projection supersedes numerical integration in the standard GFEM.

\begin{figure*}[h!]
	\setlength{\tabcolsep}{2pt}
	\centering
	\begin{tabular}{ccccc}
		\includegraphics[height=4.9cm]{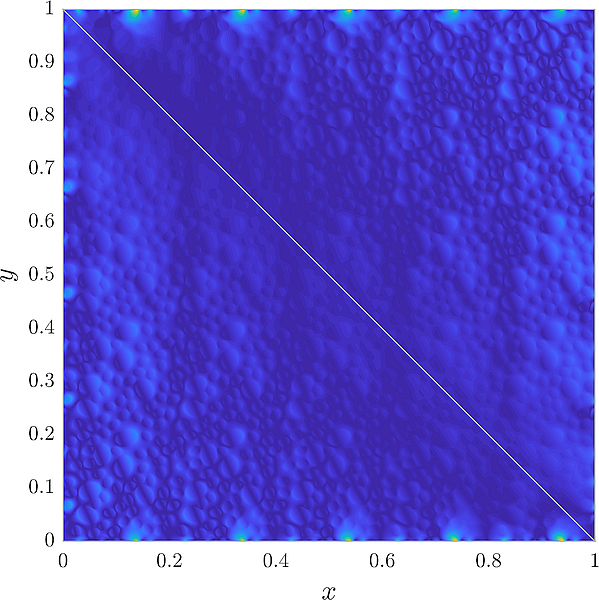} & \includegraphics[height=4.9cm]{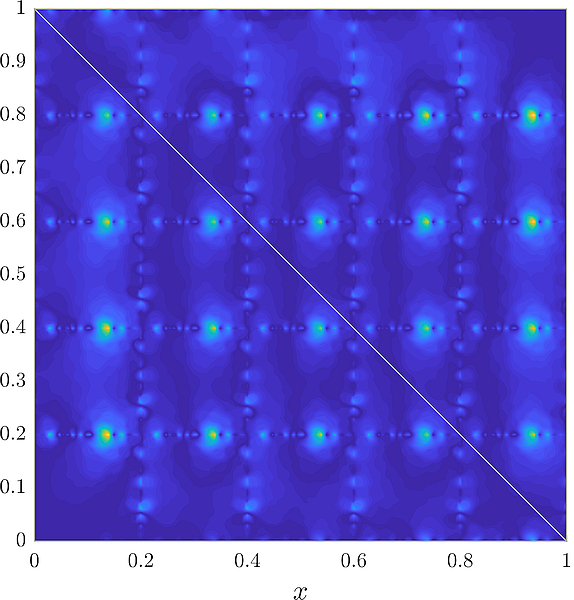} & \includegraphics[height=4.9cm]{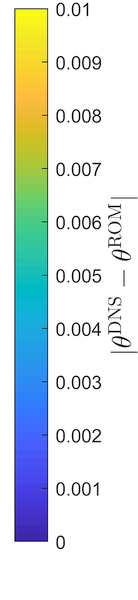} & \includegraphics[height=4.9cm]{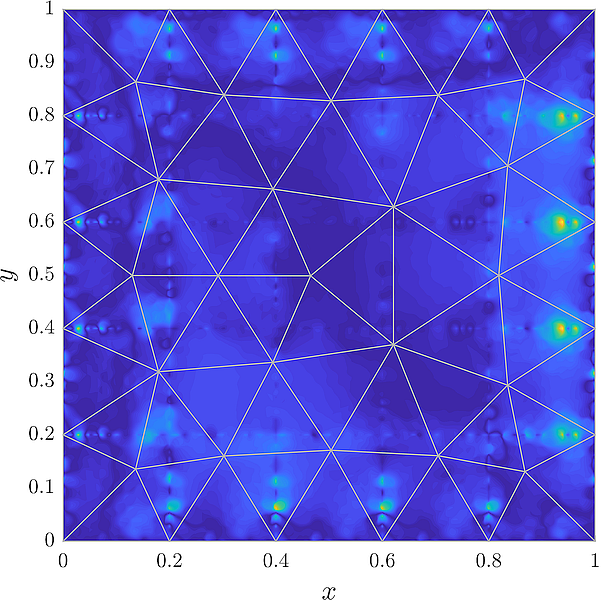} & \includegraphics[height=4.9cm]{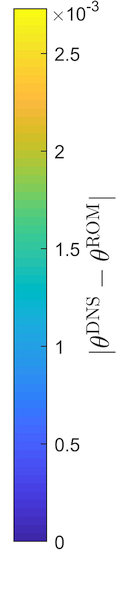} \\
		(a) & (b) & & (c) &
	\end{tabular}
	\caption{The magnitude of discrepancy between the direct numerical simulation and the result of the proposed reduced scheme for a square domain with periodic microstructure: a two-element coarse discretization (a) with only selected fluctuation fields obtained under the periodic constraint~\Eref{eq:1st_PBC} and (b) with all fluctuation fields labelled $1^{\rm{st}}$, and (c) a refined discretization with all fluctuation fields considered. Edges of the macroscopic discretization are plotted in the light grey colour.}
	\label{fig:simple_domain_discrepancy}
\end{figure*}

%
Regarding the imposition of essential boundary conditions at the macroscale, we exclude the prescribed DOFs from $\semtrx{\Phi}$. Their presence is then indirectly imposed by the action of relevant elements. Consequently, the unknown $a_i^j$'s corresponding to macroscopic elements influenced by the prescribed values are not set to zero by construction.

Note also that the projection strategy enables trivial switching to fully resolved microstructural details in parts of the macroscopic domain when desired, see~\Sref{sec:local_refinement}.

\section{Numerical examples}
\label{sec:numerical_examples}

For the numerical examples in this section, we used the two-dimensional set of 16 Wang tiles depicted in~\Fref{fig:concept_illustration}. 
Geometry of each tile was discretized using linear triangular elements. On average, each tile contained approximately 25,500 DOFs and 50,500 elements, which are refined along the inclusion boundaries.

Both material phases (depicted in light and dark grey) are considered as linear isotropic with conductivities $\tenss{K}= 10\tenss{I}$ for the matrix phase and  $\tenss{K}= 100\tenss{I}$ for the inclusion phase, respectively, where $\tenss{I}$ denotes the second-order identity tensor.

Please, note that the discussion of the results is deferred to~\Sref{sec:conclusions}.

\begin{figure*}[h!]
	\setlength{\tabcolsep}{3pt}
	\centering
	\begin{tabular}{ccc}
		(a) & \includegraphics[width=0.45\textwidth]{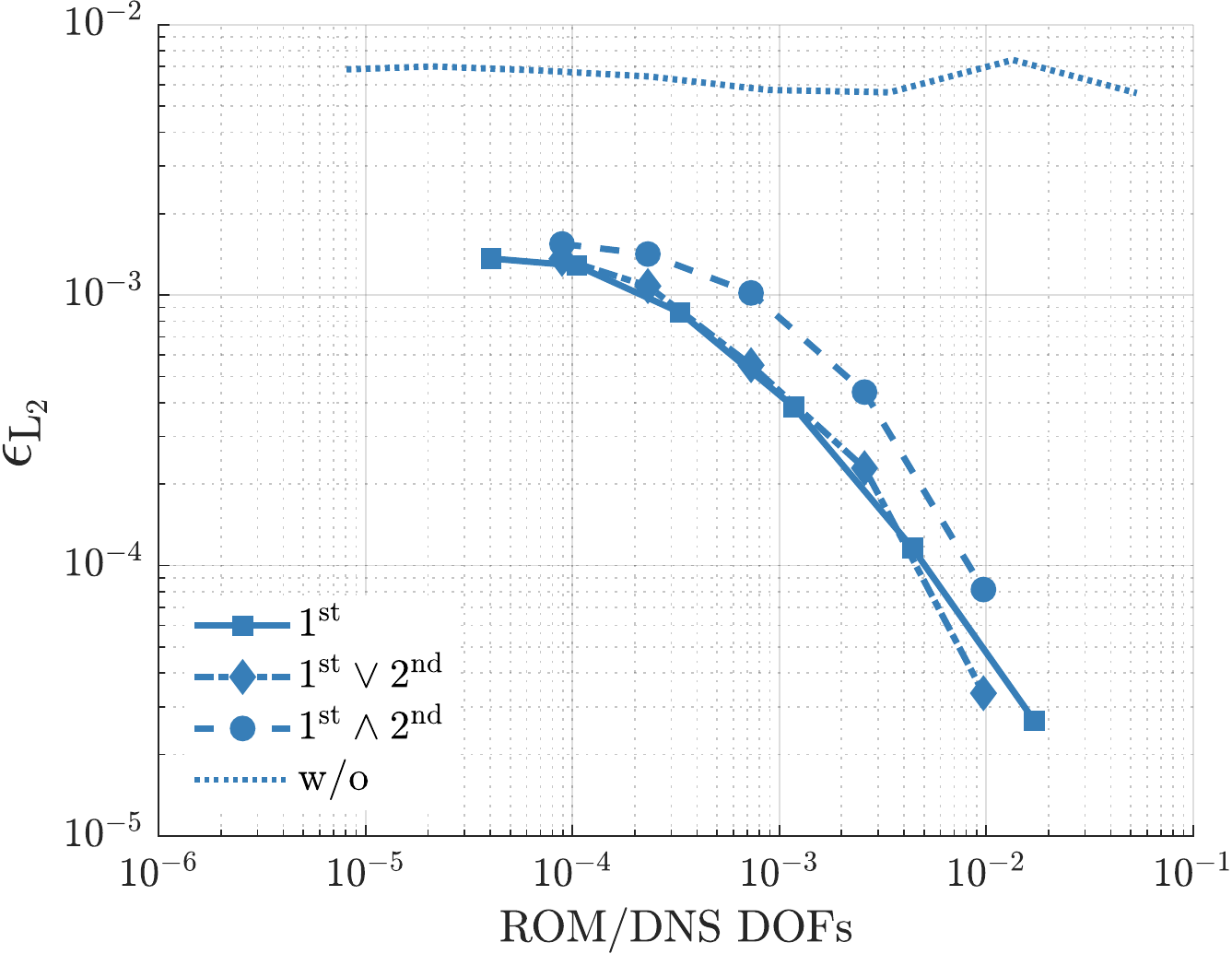} & \includegraphics[width=0.45\textwidth]{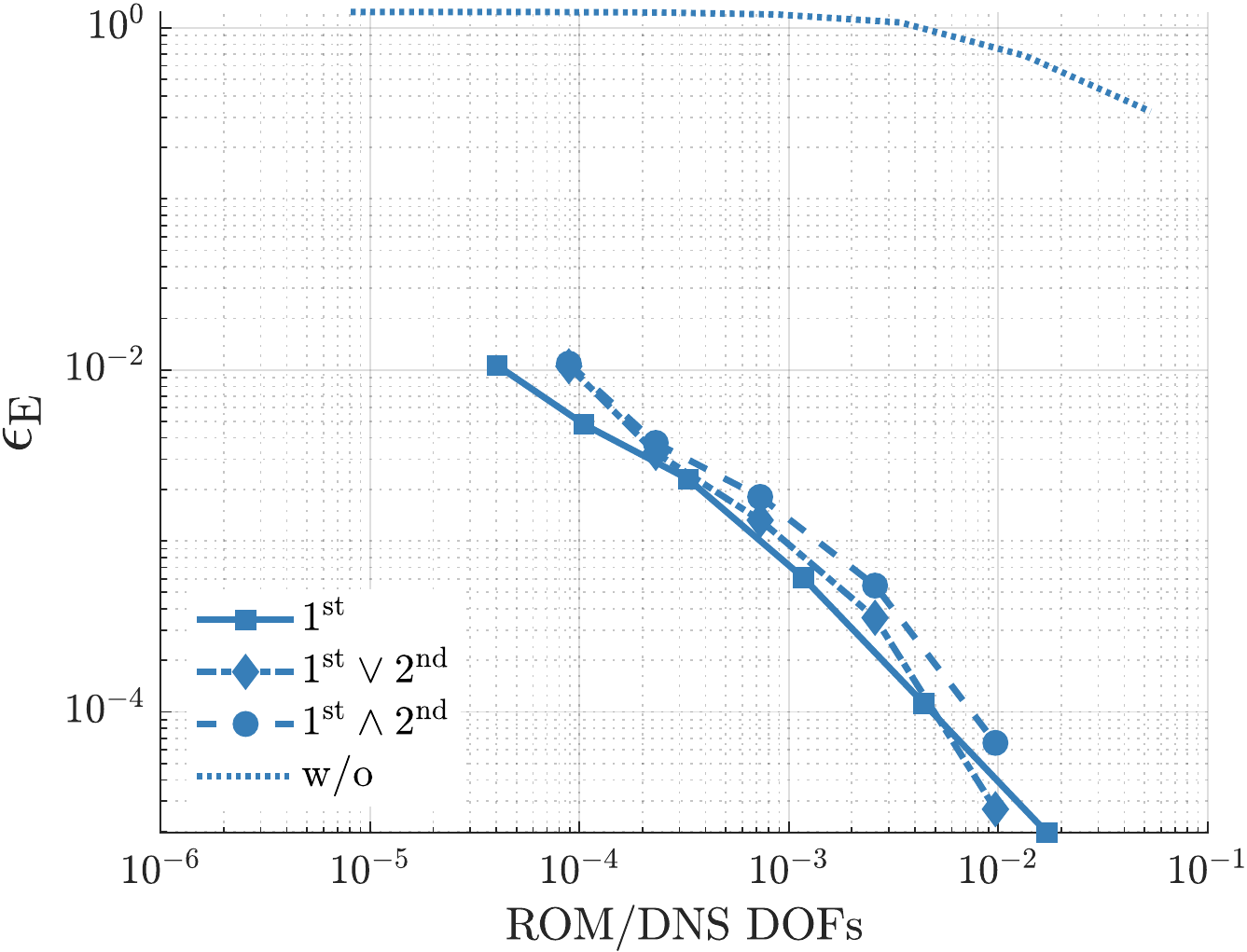} \\
		(b) & \includegraphics[width=0.45\textwidth]{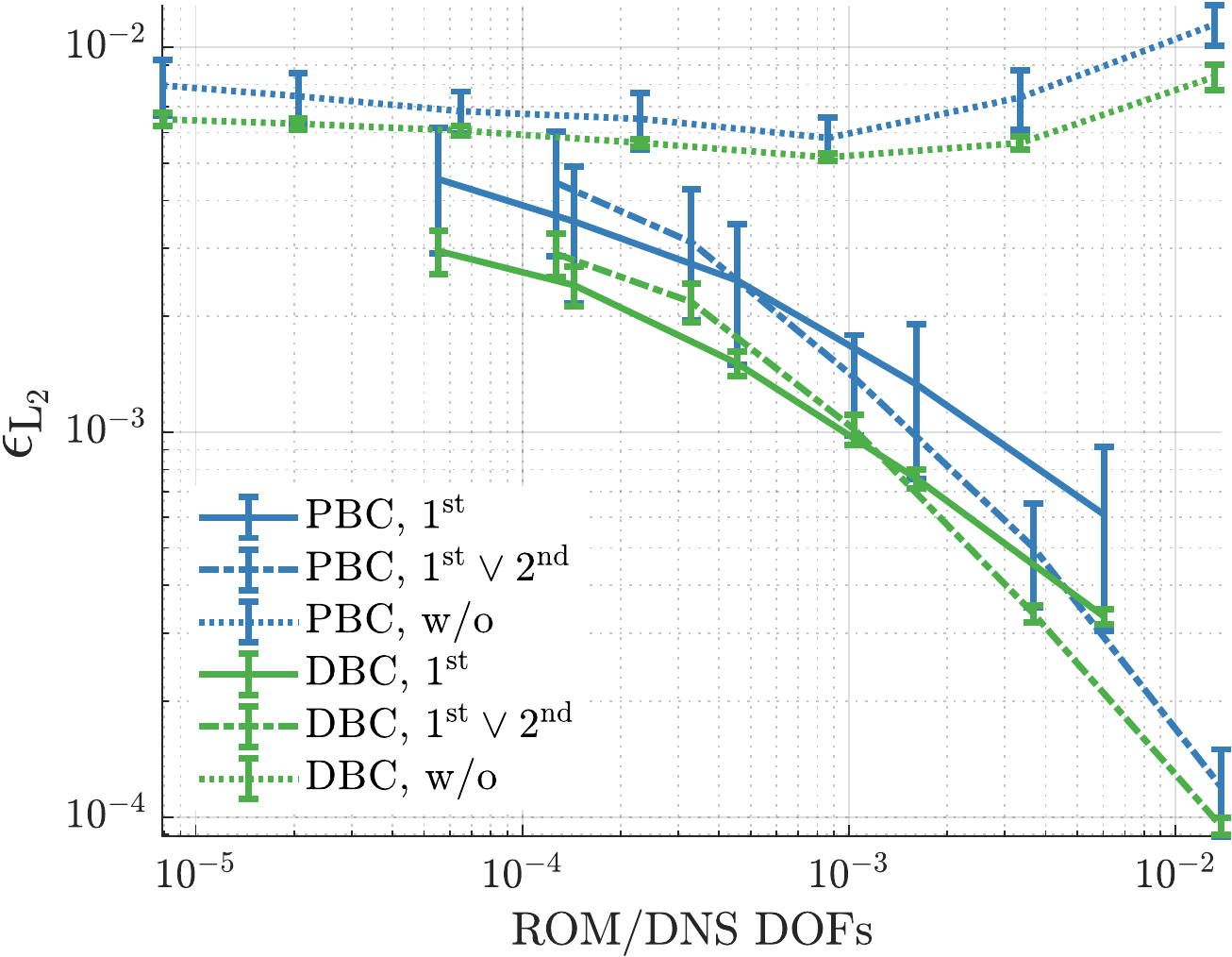} & \includegraphics[width=0.45\textwidth]{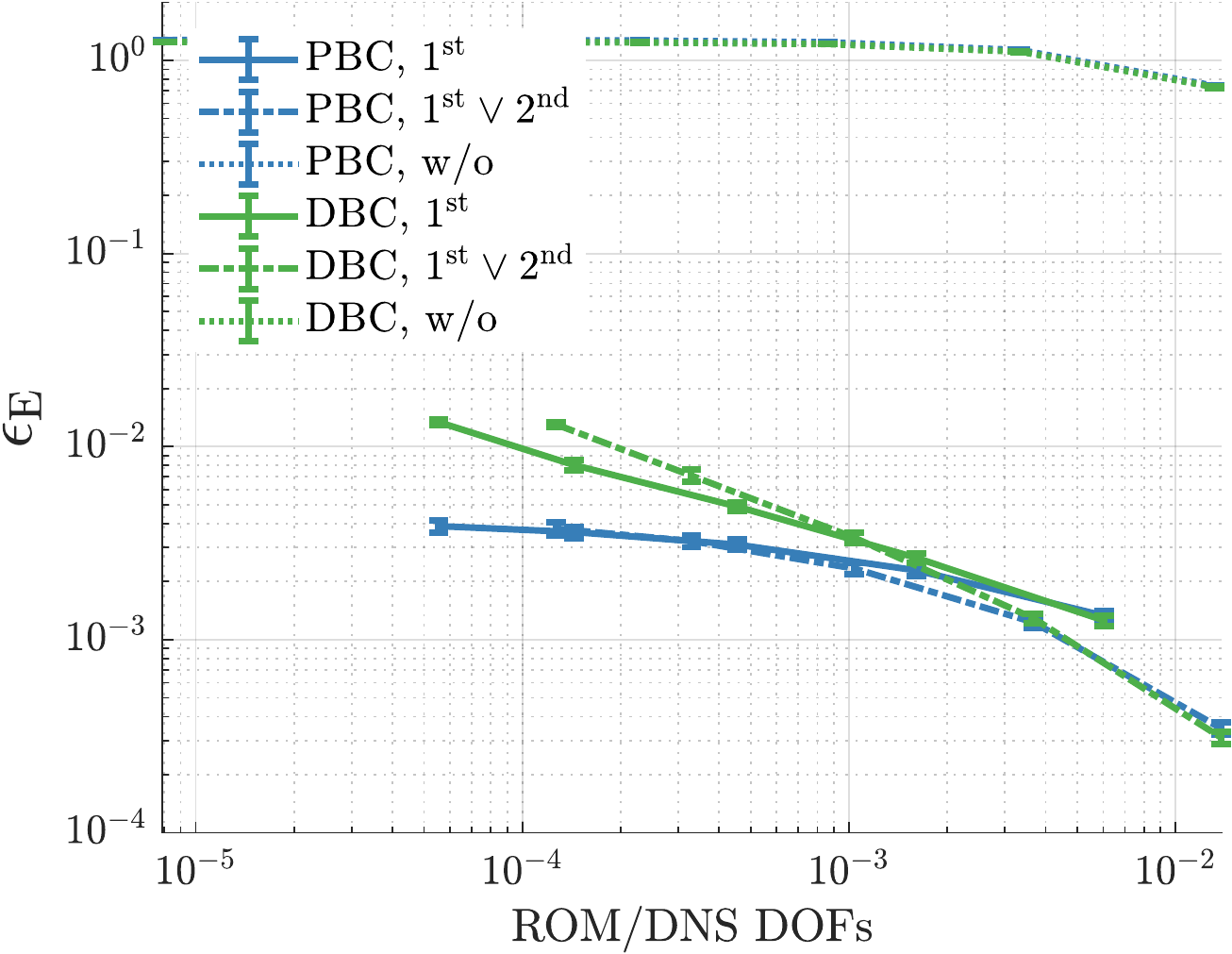} 
	\end{tabular}
	\caption{Evolution of the relative $L_2$-norm error ($\epsilon_{\text{L}_2}$) and the relative error in energies ($\epsilon_{\text{E}}$) with a uniform refinement of the macroscopic discretization for the simple square domain with (a) a periodically repeating microstructure under the Dirichlet boundary conditions and (b) a microstructure assembled from the set of 16 tiles under both periodic and Dirichlet boundary conditions. Average and the standard deviation of results obtained for 50 different realizations are plotted in (b). 
		Data labels correspond to the different choices of fluctuation fields: $1^{\text{st}}$ denotes the fields extracted with prescribing only the first-order macroscopic gradient; $1^{\text{st}}\vee2^{\text{nd}}$ and $1^{\text{st}}\wedge2^{\text{nd}}$ stand for the case when the first and second-order macroscopic gradients were prescribed independently or simultaneously, respectively. Additionally, a case when no fluctuation fields are considered (labelled w/o) is plotted for comparison.
	}
	\label{fig:simple_domain_PUC_errors}
\end{figure*}

\subsection{Uniformly loaded square sample}
\label{sec:example_square}

As the first example, we considered a simple square domain composed of five tiles in each direction with the size of the domain scaled to unity.	The domain was loaded with a uniform macroscopic gradient $\begin{bmatrix}0.6&0.3\end{bmatrix}\trn$ with boundary conditions similar to the standard first-order computational homogenization: periodic and Dirichlet ones, see~\Fref{fig:simple_domain_illustration}.

\subsubsection{PUC}

As a sanity check, we first tested the proposed framework with a trivial case of a Wang tile set---a Periodic Unit Cell.
The self-compatible tile 16 from the set shown in~\Fref{fig:concept_illustration} was chosen as the PUC and the microstructure-informed modes were recomputed considering only this tile.
Note that in the particular case of PUC, the tile-wise and set-wise constraints coincide and thus only one has to be considered in order to avoid ill-posedness of the reduced problem.

As expected, due to the linearity of the problem, the proposed reduced numerical scheme (ROM) yields exactly the same results as the Direct Numerical Simulation (DNS) even for a very coarse macroscopic discretization containing two elements, depicted in~\Fref{fig:simple_domain_discrepancy}a, when the periodic boundary conditions are assumed.
Moreover, only the fluctuation modes extracted under the periodic constraints and the first order macroscopic gradient, \Eref{eq:1st_PBC}, are sufficient in this case.
However, the same conclusions do not hold when a domain with the same periodically repeating microstructure is subjected to the macroscopic gradient under the Dirichlet boundary conditions.

In such a case, the numerical scheme only approximates the reference solution.
Figure~\ref{fig:simple_domain_discrepancy}a illustrates the discrepancy between DNS and ROM solutions with the coarsest macroscopic discretization (which was sufficient for the previous loading) and only the modes extracted under periodic constraints. 
Enriching the approximation space with fluctuation fields obtained under the Dirichlet-type constraints localizes the solution discrepancy predominantly to the tile edges; see~\Fref{fig:simple_domain_discrepancy}b.
The local difference both in its magnitude and localization can be further improved by refining the macroscopic discretization, e.g.~\Fref{fig:simple_domain_discrepancy}c. 

In addition to the local discrepancy, we quantify the error of ROM solution using (i) a relative $L_2$-norm of the results difference,
\begin{equation}
\epsilon_{\text{L}_2} = \frac{\lVert \temp^{\text{DNS}} - \temp^{\text{ROM}} \rVert_{L_2}}{\lVert \temp^{\text{DNS}}\rVert_{L_2}} \,,
\end{equation}
where
\begin{equation}
\lVert \temp \rVert_{L_2} = \left( \int_{\domain} \temp^2\atx \de{\x} \right)^{\half} \,,
\end{equation}
and (ii) a relative error in energy,
\begin{equation}
\epsilon_{\text{E}} = \frac{\lvert E^{\text{DNS}} - E^{\text{ROM}} \rvert}{E^{\text{DNS}}} \,,
\end{equation}
with 
\begin{equation}
	E^{\bullet} = \int_{\domain} \half \grad\temp^{\bullet}\atx \scontr \tenss{K}\atx \scontr \grad\temp^{\bullet}\atx \de{\x} \,.
\end{equation}

Figure~\ref{fig:simple_domain_PUC_errors}a illustrates the evolution of these errors with a uniform refinement of the macroscale discretization.
The results are shown for three choices of fluctuation fields: $1^{\text{st}}$, $1^{\text{st}}\vee2^{\text{nd}}$, and $1^{\text{st}}\wedge2^{\text{nd}}$; recall their definition in~\Sref{sec:mode_families}.
%
In addition, the results for a case when no fluctuation fields are assumed is also plotted (labelled as w/o) to highlight the influence of fluctuation fields on the errors.

\subsubsection{Set of 16 tiles}

Finally, the same analysis is repeated for a domain with a microstructure assembled from the full tile set instead of a single PUC. The global errors shown in~\Fref{fig:simple_domain_PUC_errors}b were obtained as an average over 50 stochastic realizations. Results for both the periodic and Dirichlet boundary conditions and different fluctuation field choices are reported. For the sake of brevity, we omit $1^{\text{st}}\wedge2^{\text{nd}}$ fluctuation field choice in~\Fref{fig:simple_domain_PUC_errors}b because its results were very close to those of $1^{\text{st}}\vee2^{\text{nd}}$.

\subsection{L-shape domain}
\label{sec:example_Lshape}

\begin{figure}[h!]
	\centering
	\scalebox{0.9}{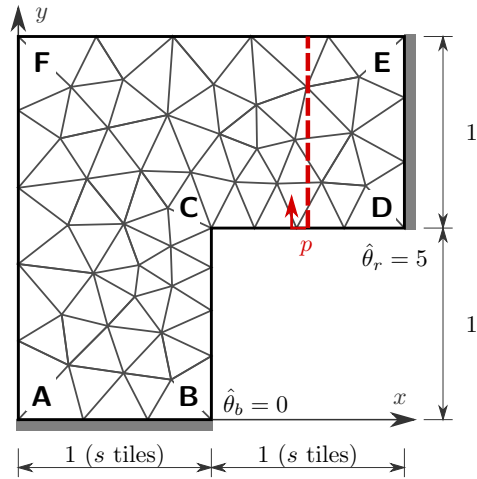}
	\caption{Illustration of L-shape domain problem. Values $\hat{\temp}$ are prescribed at the grey-shaded edges. The red dashed line (with the parametrization coordinate $p$) indicates a cross-section along which the local error profile was extracted. The triangulation depicts the initial coarse discretization for the macroscopic problem.}
	\label{fig:Lshape_definition}
\end{figure}

As the second example, we chose an L-shaped domain with values $\hat{\temp}_b=0$ and $\hat{\temp}_r=5$ prescribed at the bottom (\textsf{AB}) and right (\textsf{ED}) edge of the domain, respectively; see the illustration in~\Fref{fig:Lshape_definition}. 
The domain was scaled such that its both ends had a unit length. However, the number of tiles $s$ along each edge varied, giving a notion of scale separation of the problem.
This time, only the non-periodic microstructures assembled with Wang tilings were investigated.

\begin{figure}[h!]
	\setlength{\tabcolsep}{0pt}
	\begin{tabular}{cc}
		(a) & \includegraphics[width=0.425\textwidth]{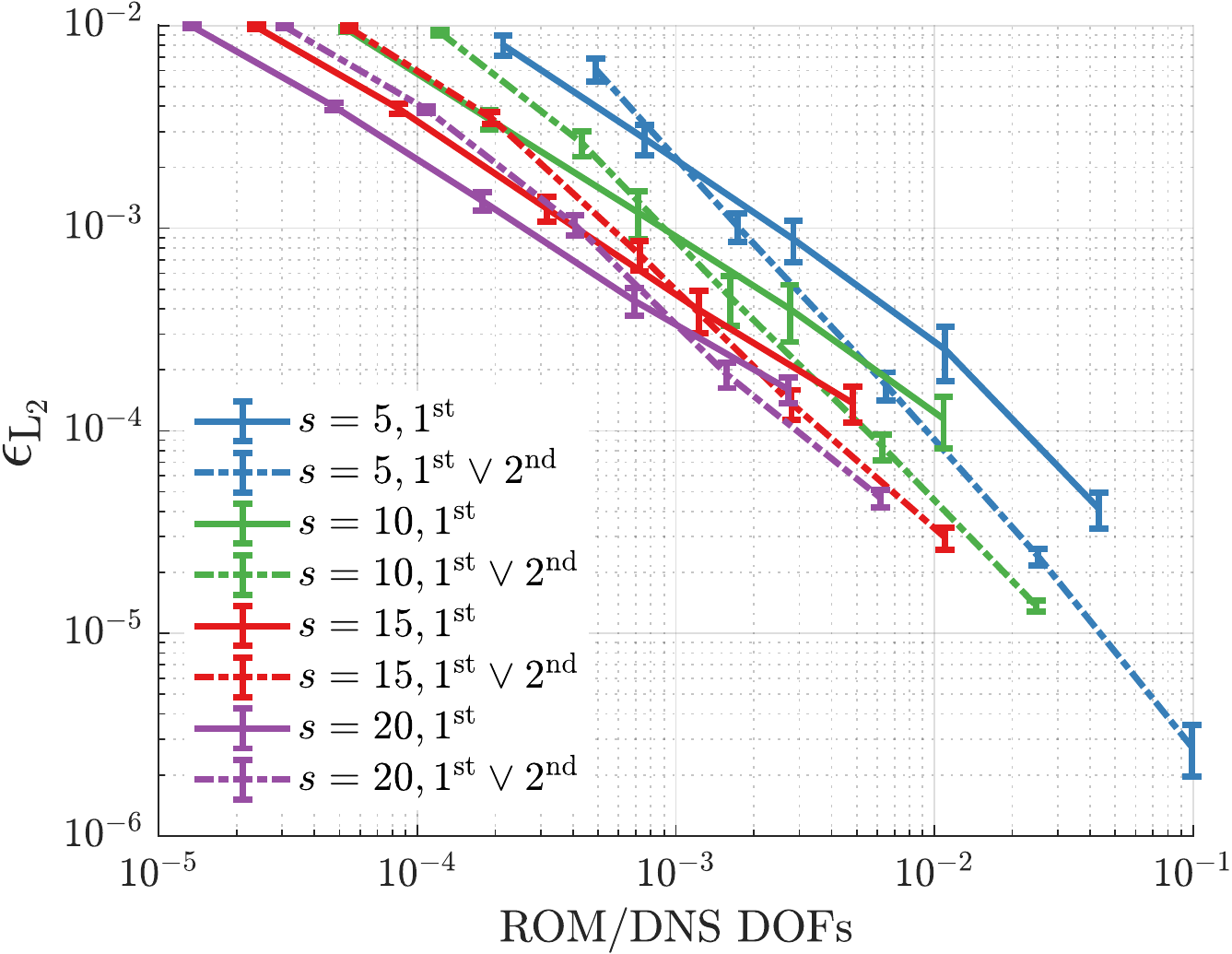} \\
		(b) & \includegraphics[width=0.425\textwidth]{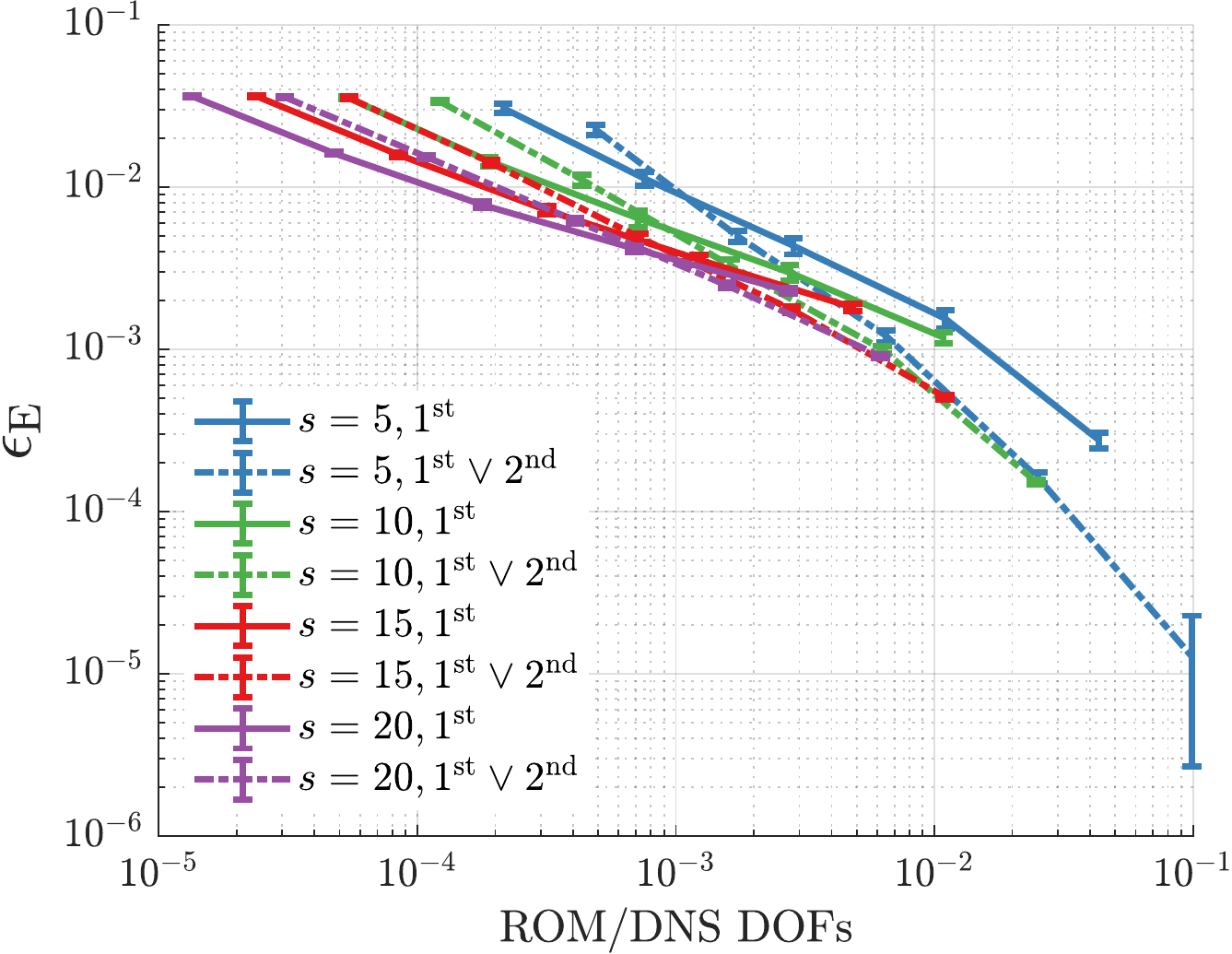} 
	\end{tabular}
	\caption{Evolution of the relative errors (a) $\epsilon_{\text{L}_2}$ and (b) $\epsilon_{\text{E}}$ in the L-shape domain problem with a uniform refinement of the macroscopic discretization and different combinations of the considered fluctuation fields and values of $s=\{5,10,15,20\}$. The average and standard deviation of the results were obtained from 10 different microstructure realizations.}
	\label{fig:Lshape_global_errors}
\end{figure}

Similarly to the previous section, we ran 10 realizations for $s\in\{5,10,15,20\}$ with four sequential refinements and different fluctuation field choices, and we recorded the global errors $\epsilon_{\text{L}_2}$ and $\epsilon_{\text{E}}$; see~\Fref{fig:Lshape_global_errors}. Similarly to the results in~\Fref{fig:simple_domain_PUC_errors}b, only the $1^{\text{st}}$ and $1^{\text{st}}\vee2^{\text{nd}}$ fluctuation field choices are eventually shown, because the results of $1^{\text{st}}\vee2^{\text{nd}}$ and $1^{\text{st}}\wedge2^{\text{nd}}$ were almost identical.

\begin{figure}[h!]
	\setlength{\tabcolsep}{0pt}
	\centering
	\begin{tabular}{c}
		\includegraphics[width=0.45\textwidth]{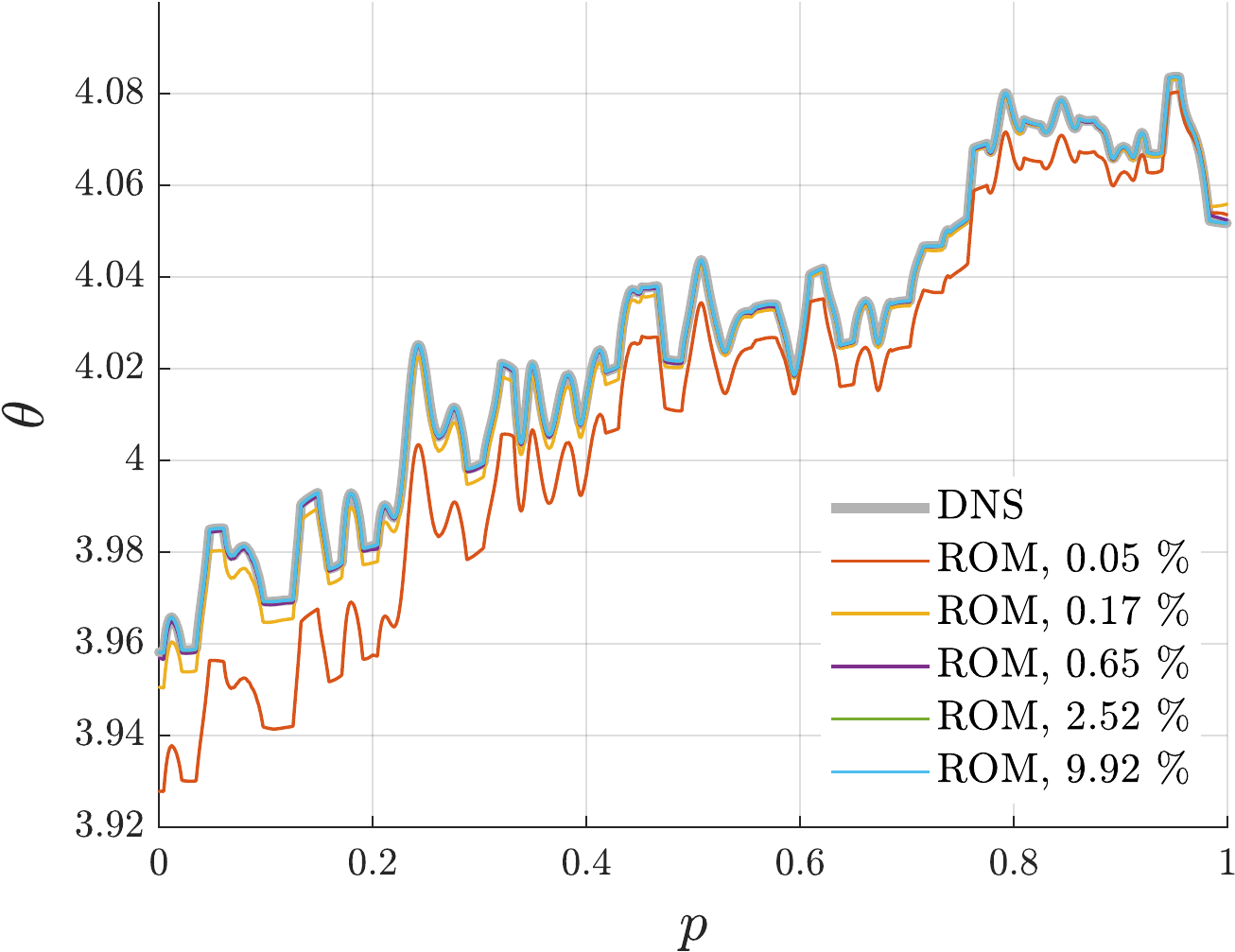}
	\end{tabular}
	\caption{The point-wise convergence of a solution profile taken along the cross-section depicted in~\Fref{fig:Lshape_definition} with a dashed red line and parametrized with a relative coordinate $p\in\interval{0.0}{1.0}$. All results are shown for $s=5$. Different line colours correspond to consecutive refinements of the initial macroscopic discretization (the legend also states the ratio between the number of ROM and DNS degrees of freedom).}	
	\label{fig:Lshape_local_errors}
\end{figure}

Figure~\ref{fig:Lshape_local_errors} illustrates the point-wise convergence for one microstructure realization for $s=5$ with a solution profile along the cross-section depicted in~\Fref{fig:Lshape_definition}. The cross-section was chosen such that it cuts through the centres of tiles. However, similar trends were observed also in the cross-section coinciding with the tile edges.

\subsubsection{Local refinement}
\label{sec:local_refinement}

\begin{figure*}
	\centering
	\begin{tabular}{cccc}
		\includegraphics[height=4.5cm]{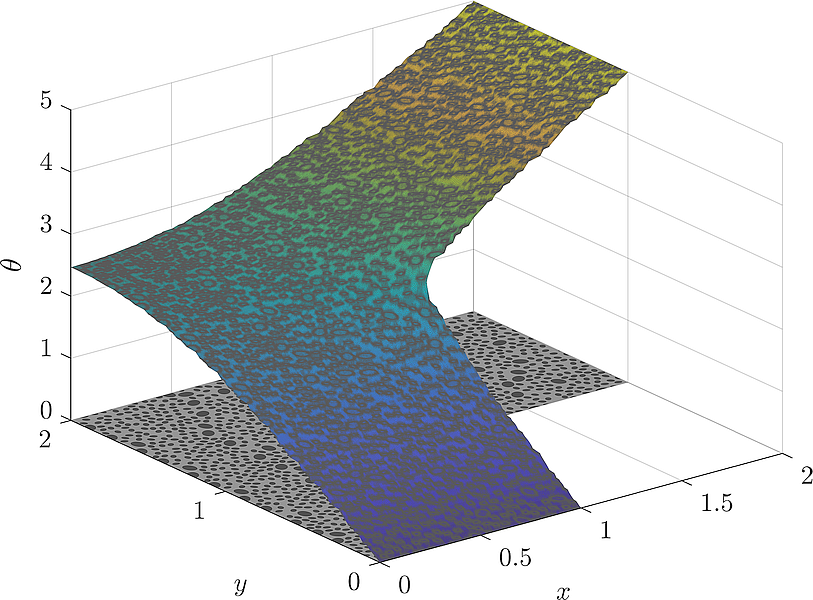} & \includegraphics[height=4.5cm]{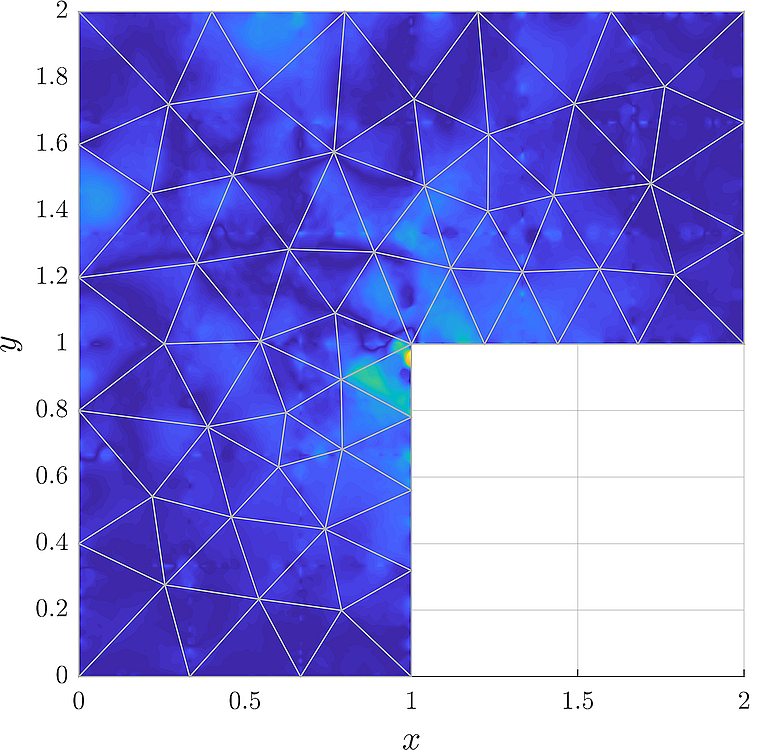} & \includegraphics[height=4.5cm]{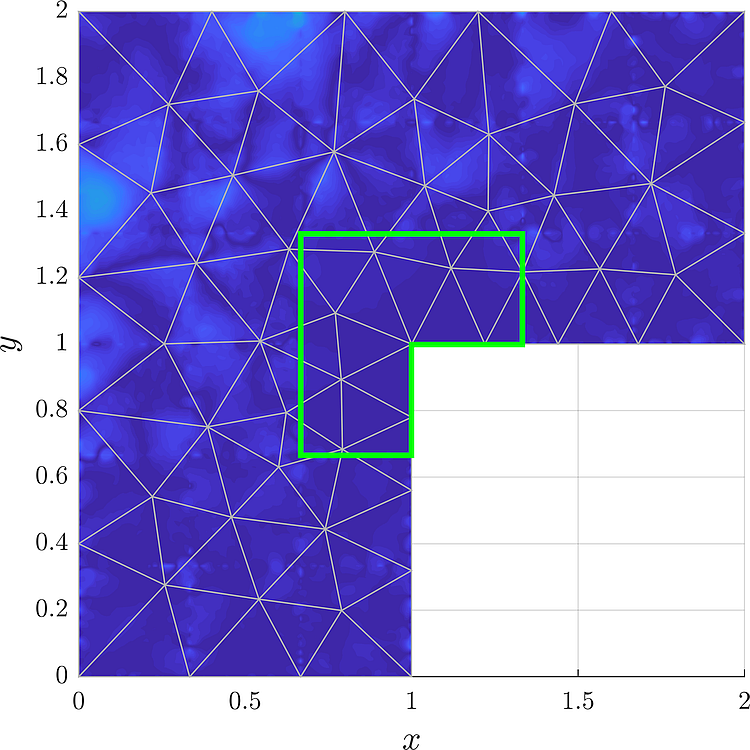} & \includegraphics[height=4.5cm]{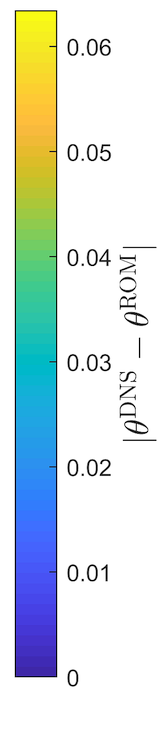} \\
		(a) & (b) & (c) &
	\end{tabular}
	\caption{An example of (a) a DNS solution for $s=3$ and the local errors of the reduced scheme without (b) and with (c) the microstructure fully resolved in the highlighted region.}
	\label{fig:local_refinement}
\end{figure*}

As mentioned in the previous section, adopting the viewpoint of Reduced Order Modelling with GFEM as a mean of constructing the final reduced modes allows for straightforward switching to a discretization resolving all microstructural details in selected regions. As an illustration, \Fref{fig:local_refinement}a depicts a solution to the L-shape problem for one realization with $s=3$. The plot of local differences, \Fref{fig:local_refinement}b, reveals that the largest local errors are expectedly concentrated around the re-entrant corner.

Thus, a region containing 3 tiles around the corner was marked for the local refinement and the analysis was repeated. Local errors of this refined model are shown in~\Fref{fig:local_refinement}c. Global errors reduced from $2.654\times10^{-3}$ to $1.393\times10^{-3}$ for $\epsilon_{\text{L}_2}$, and from $1.055\times10^{-2}$ to $5.712\times10^{-3}$ for $\epsilon_{\text{E}}$, respectively. Note that, given the small separation of scales (only 27 tile domains were present in the domain altogether), the local refinement leads to a significant increase in unknowns from $0.14\%$ to $11.16\%$ of all DOFs present in the domain.

\subsubsection{Comparison against ROM}
\label{sec:standard_ROM}

\begin{figure*}
	\setlength{\tabcolsep}{0pt}
	\centering
	\begin{tabular}{ccc}
		\includegraphics[height=4.4cm]{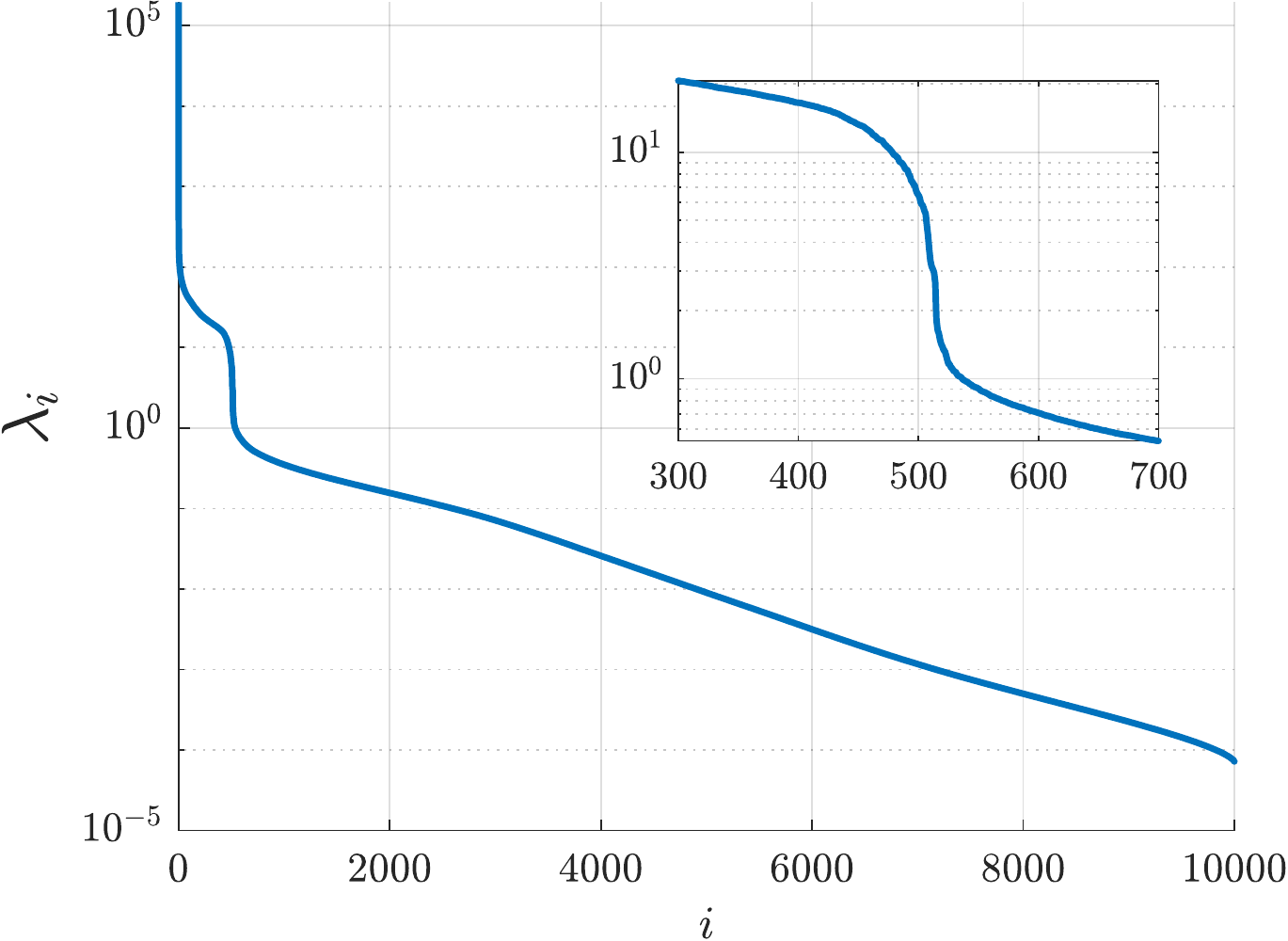} & \includegraphics[height=4.5cm]{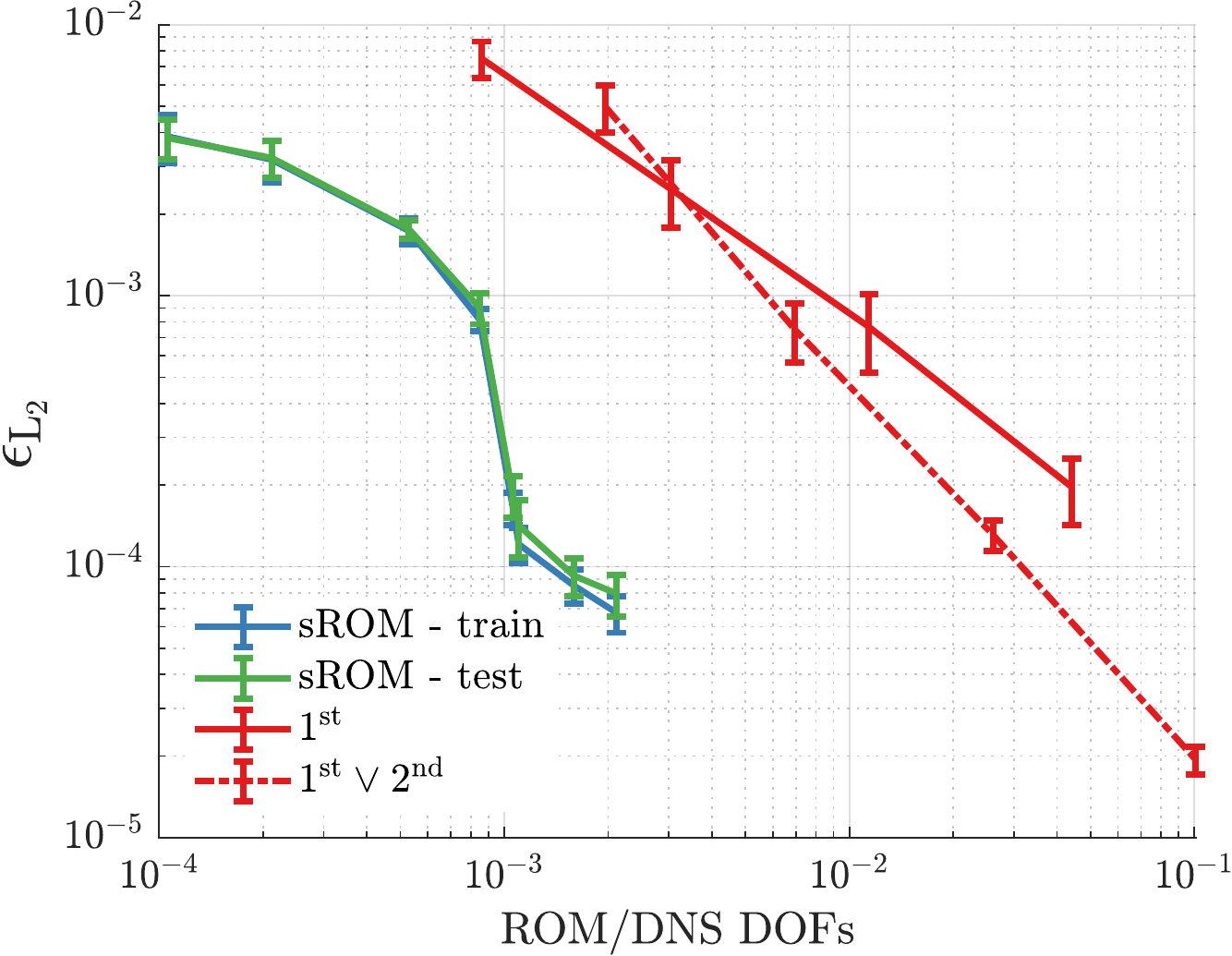} & \includegraphics[height=4.5cm]{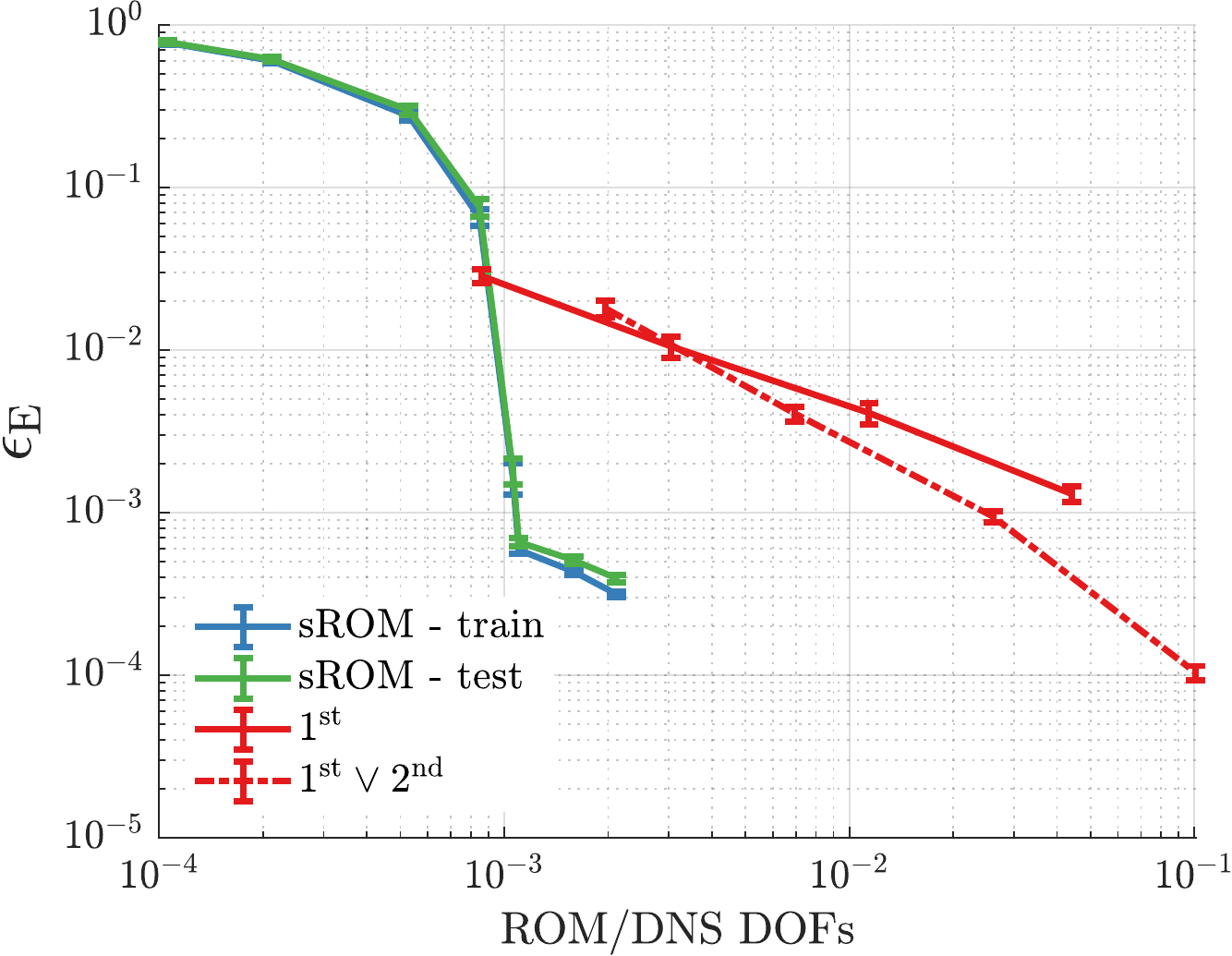} \\
		(a) & (b) & (c)
	\end{tabular}
	\caption{Results of the snapshot-based Reduced Order Modelling (labelled as sROM) scheme described in \Sref{sec:standard_ROM} for the L-shape domain example with $s=4$: (a) eigenvalues from Singular Value Decomposition of 10,000 collected snapshots, and the relative (b) $L_{2}$-norm error $\epsilon_{\text{L}_2}$ and (c) error $\epsilon_{\text{E}}$ in energies computed for 50 unique microstructure realizations both contained in the snapshots (in blue) and not used in the training phase (in green), compared to the proposed microstructure-informed reduced scheme (in red).}
	\label{fig:POD_errors}
\end{figure*}

In order to put the number of necessary modes into perspective of the classical snapshot-based Reduced Order Modelling strategy, we devise the following test:
While keeping the same microstructural geometry as the one shown in~\Fref{fig:concept_illustration}, we replace the unstructured finite element discretization with a regular, pixel-based mesh, in which two triangular elements with alternating orientation constitute one pixel. Each tile is then represented with $100\times100$ nodes and 19,602 finite elements.
Consequently, each microstructural realization features the same discretization and hence solutions of the illustrative L-shape domain example can be straightforwardly stored as a snapshot without the need for a remapping/projection onto a fixed mesh.

We generate 10,000 distinct microstructural realizations for the L-shape domain with $s=4$ and solve the temperature field for each.
We then perform Singular Value Decomposition for the matrix containing individual snapshots as its columns; \Fref{fig:POD_errors}a displays the eigenvalue magnitudes.
Next, we replace the GFEM-based microstructure-informed modes $\semtrx{\Psi}$ in the analysis with $n$ eigenmodes corresponding to $n$ largest eigenvalues and compute the response of the compressed system for $n\in\left\{50,100,250,400,500,550,1000\right\}$ and plot the global errors $\epsilon_{\text{L}_2}$ and $\epsilon_{\text{E}}$ in Figures \ref{fig:POD_errors}a and \ref{fig:POD_errors}b, respectively. The errors of the snapshot-based ROM are also compared with the scheme proposed in this work applied to the above mentioned discretization.

\begin{figure*}
	\setlength{\tabcolsep}{0pt}
	\centering
	\begin{tabular}{cc}
		\includegraphics[height=4.4cm]{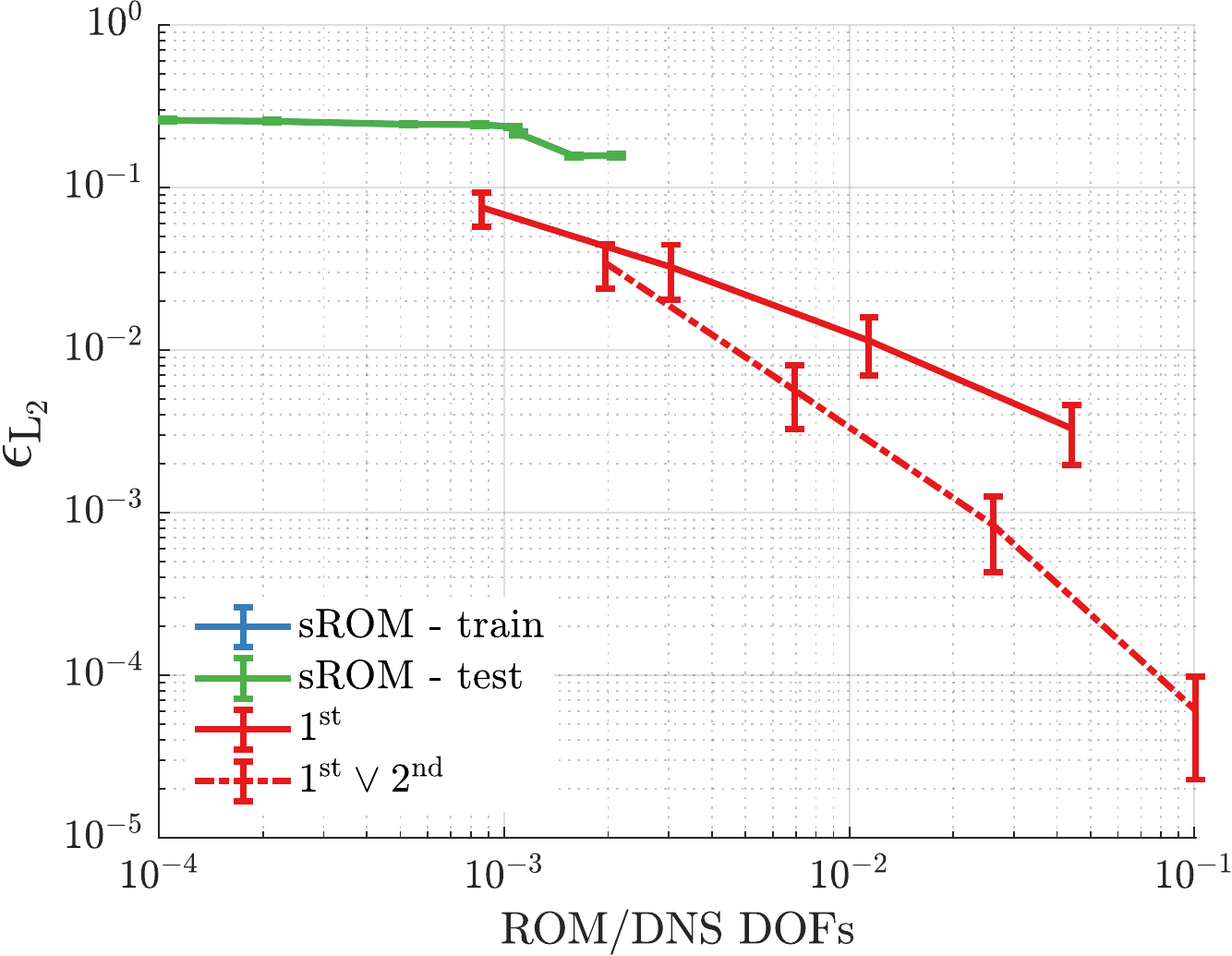} & \includegraphics[height=4.5cm]{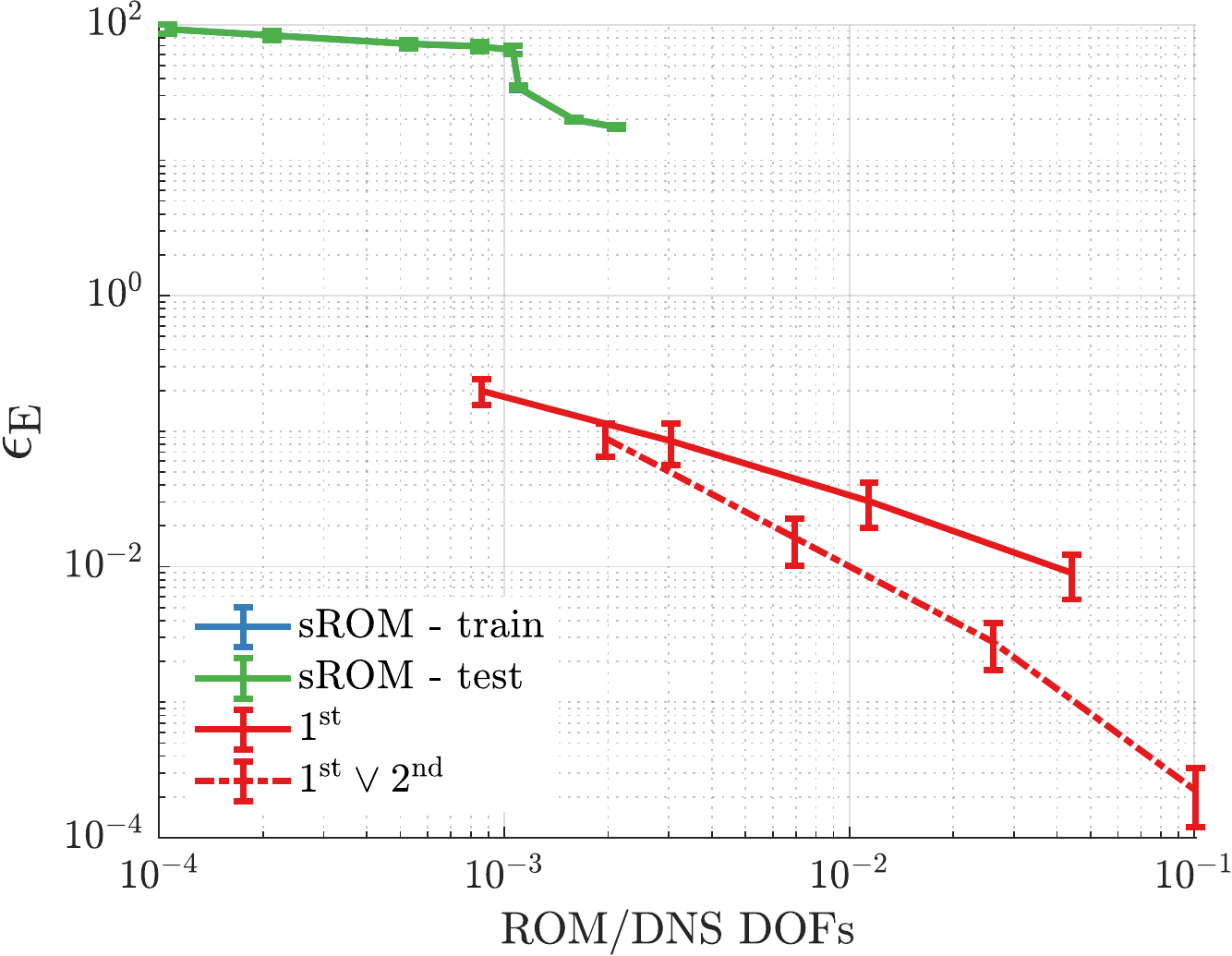} \\
		(a) & (b)
	\end{tabular}
	\caption{Comparison of the standard snapshot-based Reduced Order Modelling strategy (sROM) and the proposed microstructure-informed reduced scheme (in red) in terms of (a) $L_{2}$-norm error $\epsilon_{\text{L}_2}$ and (b) error $\epsilon_{\text{E}}$ in energies computed for 50 unique microstructure realizations with the \emph{modified loading conditions}. The basis for sROM was computed from the snapshots obtained under the \emph{original loading conditions}. The blue lines correspond to errors related to microstructural realizations considered during snapshot extraction, while the green lines depict results of realizations not contained in the original training phase.}
	\label{fig:POD_mod_errors}
\end{figure*}

Finally, keeping the same geometrical setup, we also consider modified loading conditions such that zero temperature is fixed at the inner vertical edge (\textsf{BC}) and $\hat{\temp}=5$ is prescribed at the top edge (\textsf{FE}). To illustrate the universality of our GFEM-based modes when compared to the standard snapshot-based ROM, we test reusing the snapshot-based modes from the previous loading conditions and report the corresponding errors in~\Fref{fig:POD_mod_errors}.

\subsubsection{Time comparison}

%
Even though comparing the wall-clock time is always biased by the implementation, chosen solvers, and the hardware configuration, we provide calculation times of our implementation (limited to the setup described below) for the sake of completeness.

Originally, the whole framework was developed and tested using an in-house \Matlab{} implementation. However, similarly to other interpreted languages, \Matlab{} is inefficient with low-level raw loops, which should be replaced with a suitable vectorization whenever possible. Unfortunately, the construction of the reduced basis $\semtrx{\Phi}$ cannot be efficiently vectorized in \Matlab{}.
In order to provide a fair comparison, solvers for both DNS and ROM were eventually implemented as a C/C++ MEX functions\footnote{\href{https://www.mathworks.com/help/matlab/call-mex-file-functions.html}{https://www.mathworks.com/help/matlab/call-mex-file-functions.html}}, using Eigen~\cite{eigenweb} as a C++ library for linear algebra.
Moreover, the construction of the stiffness matrix $\semtrx{K}$ and the reduced basis $\semtrx{\Phi}$ was parallelized using OpenMP interface.

\begin{figure}
	\centering
	\includegraphics[width=0.45\textwidth]{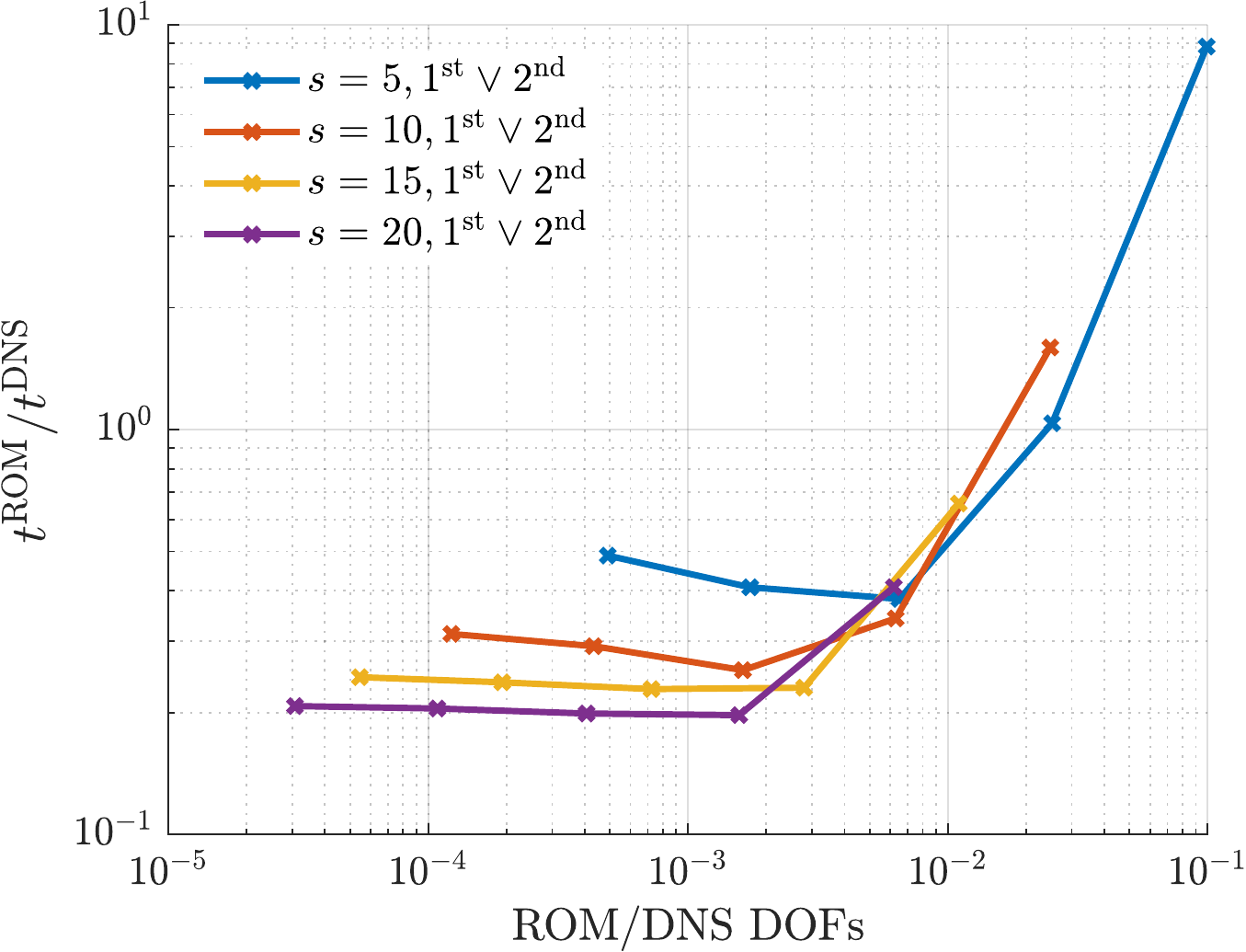}
	\caption{Average computational time of a reduced problem relative to DNS measured for 10 different microstructural realizations. All reported times were obtained using a workstation equipped with an Intel\textsuperscript{\textregistered} Xeon\texttrademark{} Gold 6144 3.50 GHz processor and 256 GB RAM.}
	\label{fig:Lshape_timing}
\end{figure}

Figure~\ref{fig:Lshape_timing} shows the average computational time needed for reduced problems relative to the time needed to solve DNS. The refinement levels correspond to the errors given in Figures \ref{fig:Lshape_global_errors} and \ref{fig:Lshape_local_errors}.
An iterative solver with preconditioning was employed for DNS while ROM-related linear system of equations were solved using a direct sparse solver based on the Cholesky decomposition with pivoting.
%
%

\section{Conclusions}
\label{sec:conclusions}

Incorporating the pre-computed fluctuation fields as a basis for constructing the microstructure-informed reduced modes enables analyses of models with fully resolved microstructural details with a fraction of degrees of freedom compared to the fully detailed model.
The comparison against the coarse discretization without enrichments, recall \Fref{fig:simple_domain_PUC_errors}, clearly demonstrates that a similar reduction in unknowns cannot be achieved without modes which efficiently approximate the local character of the solution.
Even for the coarsest discretizations with four orders of magnitude fewer DOFs than the fully resolved models, both the relative global errors $\epsilon_{\text{L}_2}$ and $\epsilon_{\text{E}}$ were already below 3\% in all examples.

%
Higher accuracy can be achieved in two ways: (i) by considering more fluctuation fields when constructing the microstructure-informed reduced modes, recall~\Eref{eq:GFEM_ansatz}, or (ii) by refining the coarse discretization of the macroscopic domain.
Irrespectively whether the domain is loaded in such a way it exhibits a constant (\Sref{sec:example_square}) or varying (\Sref{sec:example_Lshape}) macroscopic gradient, the benefits of refining the discretization while assuming only $1^{\text{st}}$ fluctuation fields prevail only to a certain threshold. 
With the size of macroscopic elements below the size of a tile, adding the fluctuation fields obtained also for the second-order macroscopic gradient results in lower errors and better convergence rate. Interestingly, the convergence rate of $\epsilon_{\text{L}_2}$ was almost identical for $1^{\text{st}}$ case in all tested values of $s$, while it exhibited dependence on $s$ in $1^{\text{st}}\vee2^{\text{nd}}$ case of $\epsilon_{\text{L}_2}$ and in both cases of $\epsilon_{\text{E}}$.

Except for the simplest example of a uniformly loaded domain with periodic microstructure, we also observed only a minor difference between the fluctuation fields obtained for the first- and second-order macroscopic gradient prescribed separately ($1^{\text{st}}\vee2^{\text{nd}}$) or simultaneously ($1^{\text{st}}\wedge2^{\text{nd}}$), with the latter option slightly outperforming the former for refined macroscopic discretizations in terms of solution errors.

%
Local discrepancy between the reduced models and DNS appears in two forms. First, more uniformly distributed discrepancy arises from an insufficiently coarse macroscopic discretization unable to capture the overall kinematics of the problem, see~\Fref{fig:local_refinement}b. 
The straightforward remedy here is to refine the dicretization.
For instance, the local solution profile in~\Fref{fig:Lshape_local_errors} improves significantly already upon the first refinement; the second refinement then yields the solution of the reduced scheme that is almost indistinguishable from DNS.

The second form of discrepancy reflects the underlying microstructure and stems from an inaccurate approximation by the pre-computed fluctuation fields; compare e.g. Figures \ref{fig:simple_domain_discrepancy}a and \ref{fig:simple_domain_discrepancy}b.
Taking into account more fluctuation fields improves these local errors. Yet, because the number of fluctuation fields is limited, some errors will always linger (albeit with a diminishing magnitude). These errors localize mainly along tile boundaries, where the effect of the modelling assumptions during the extraction of the fluctuation fields is the most pronounced; see for instance the local discrepancies at tile edges next but parallel to the domain boundary in \Fref{fig:simple_domain_discrepancy}c.

Finally, the solution of the reduced model can be further enhanced by resorting to the fine, microstructure-resolving discretization in selected parts of the macroscopic domain as demonstrated in~\Sref{sec:local_refinement}; see also~\Fref{fig:local_refinement}. 

%
Admittedly, the proposed microstructure-informed reduced scheme requires more degrees of freedom than the standard snapshot-based ROM, recall \Sref{sec:standard_ROM} and \Fref{fig:POD_errors}. The reported reduction in degrees of freedom thus may seem unsatisfactory given the fact that we are dealing with a linear, scalar problem. 
However, as follows from \Fref{fig:POD_mod_errors}, any significant change in macroscopic loading necessitates recalculation of the basis for the standard ROM.
Results from \Sref{sec:standard_ROM} show that problems with fully resolved stochastic microstructures are inherently high dimensional, despite the linearity of the considered constitutive material law. Even a very small problem with $s=4$ and a simplified finite element discretization requires $\approx500$ modes for an acceptable accuracy.
Thus, even in the most favourable case, at least $500$ simulations with fully resolved microstructures must be performed in order to extract a relevant basis for the snapshots-based Reduced Order Modelling scheme. Note that for the vertex-based tile set considered throughout this work, the L-shape problem with $s=4$ allows for $2^{65}\approx3\times10^{19}$ unique microstructure realizations; however, the drop threshold in the number of required modes seems to be proportional to the number of tile positions multiplied by tile types, i.e. $48\times16=768$ in this particular case.
By contrast, our fluctuation fields are pre-computed only once at the level of tiles and they can be re-used for different macroscopic analyses when transformed into the microstructure-informed reduced modes using the GFEM ansatz.

Following from our tentative comparison, the computational time of our reduced scheme depends on the macroscopic discretization and the size of the original problem.
In general, the reduction in computational cost is not directly proportional to the reduction in the number of unknowns.
For the coarsest macroscopic mesh, the reduced scheme takes from a half to one fifth of the time needed to solve DNS, with a more favourable comparison holding for larger problems.
Time savings upon refinement follow from the interplay between the assembly of the reduced system, including construction of the reduced modes and the subsequent Galerkin projection, and solution of the reduced system of linear equations, which inherits the sparsity pattern of the coarse discretization. Eventually, there is a threshold at which the time saved by solving a smaller system cannot compensate for the time needed to construct the system.
Albeit the time comparison is pertinent to the adopted Galerkin projection at the global matrix level, i.e. following exactly \Eref{eq:reduced_system}, and other implementations are possible, similar observations are a common problem of all reduced order modelling strategies, and additional accelerations such as Hyper-reduction~\cite{ryckelynck_hyper-reduction_2009} or Discrete Empirical Interpolation Method~\cite{chaturantabut_nonlinear_2010} should be incorporated to further improve their performance.

\section{Summary}
\label{sec:summary}

%
Corroborating the potential the Wang tiling concept has in modelling heterogeneous materials with stochastic microstructure, we have introduced a numerical scheme which reduces the number of unknowns in macroscopic analyses with underlying microstructural geometry generated from a set of Wang tiles.

Unlike our previous work~\cite{novak_compressing_2012}, which approximates stress fields in heterogeneous materials and must be directly incorporated in creating the compressed microstructural representation, we have developed a procedure that extracts primal-unknown fluctuation fields non-intrusively, and can be therefore used with any existing Wang-tile based representation.
Inspired by computational homogenization, we extract the characteristic fluctuation fields as collective responses of individual Wang tiles to a parametrized loading, represented in our case with uniform and linear macroscopic gradients. 

Since the extracted fields defined at the tile level are by construction continuous across edges with the same code, they can be assembled within a macroscopic domain in the same manner as the microstructural geometry, creating an approximation for the fluctuation part of the solution providing the macroscopic domain is homogeneously loaded.
These approximations are finally transformed into a microstructure-informed reduced modes using the Generalized Finite Element Method's ansatz in order to locally interpolate between individual approximations and thus to adapt to a non-uniform macroscopic loading.

%
We have illustrated the proposed approach with two numerical examples. Even with a relatively coarse discretization (containing less than 0.01\% of the unknowns in the unreduced problem), we were able to achieve less then 3\% error in the relative $L_2$-norm and energy difference.
We demonstrated that the accuracy is further improved by (i) refining the macroscopic discretization and (ii) considering more pre-computed fluctuation fields. Moreover, the proposed numerical scheme allows for a simple transition to a fully-resolved/high-fidelity model in regions of interest.

%
Admittedly, compared to the traditional, snapshot-based Reduced Order Modelling scheme, our approach requires more modes for a comparable accuracy. This stems from the fact that, unlike the standard applications where the modes are obtained from a very similar macroscopic problems, our modes are pre-computed without any information on the macroscopic geometry or loading.
However, we have demonstrated that problems with a fully resolved stochastic microstructure inherently possess a high dimensional solution manifold and, consequently, a high number of training problems must be solved in order to extract accurate basis when the loading changes significantly.
Conversely, our pre-computed fields are universal and can be used in various macroscopic analyses.


%
Even though the proposed methodology for extracting the characteristic responses and reusing them in macroscopic analyses was illustrated with the simplest case of a two-dimensional, scalar problem, it can be straightforwardly applied to elasticity and three-dimensional applications as well.
This only requires replacing the particular form of constraints (Eqs. (\ref{eq:1st_DBC})--(\ref{eq:2nd_NBC})) with a form related to a chosen problem and/or considering the corresponding unit load cases.

%
In principle, the same procedure can be also used for a non-linear material response. In such a case, the extraction of fluctuations fields for the unit load cases is insufficient. Instead, the space of the parametrized macroscopic loading during the fields' extraction must be carefully explored, e.g.~\cite{goury_automatised_2016}, and the pre-computed fields stored and post-processed using, for instance, Proper Orthogonal Decomposition approach~\cite{kerschen_method_2005}.
However, applicability of such an approach depends on the number of resulting fields, since too many fields needed to capture the characteristic fluctuations in microstructural response render the reduced model inefficient.
%
%
Covering material non-linearities within the proposed framework constitutes the next step in our research.

%
Related to the non-linearities is also a question of numerical efficiency. 
Similarly to other reduced order modelling schemes, reducing the number of unknowns in our strategy does not translate directly to the same reduction in computational times, which is critical in non-linear problems.
As a remedy, Hyper-reduction~\cite{ryckelynck_hyper-reduction_2009}, Discrete Empirical Interpolation Method~\cite{chaturantabut_nonlinear_2010}, and many other methods have been developed to approximate the Galerkin projection in~\Eref{eq:reduced_system} with only a subset of unknowns/quadrature points; see e.g. a brief overview in~\cite{fritzen_algorithmic_2018}.
Our preliminary tests with Hyper-reduction indicate that modifications of these methods are necessary for our microstructure-informed modes, because these methods have been developed primarily for modes with global support. However, due to the GFEM-based construction, our modes have a local support inherited from the macroscopic discretization. Consequently, the standard strategies for choosing the suitable subset of unknowns do not work in our case.
Therefore, efficient implementation and the right choice and modifications of the acceleration methods remains an open question to be investigated.

\section*{Acknowledgment}
This research was funded by the Czech Science Foundation, project No. 19-26143X.
Martin Doškář also gratefully acknowledges the support from Fulbright Commission Czech Republic that funded his research stay at the University of California, San Diego.

\bibliography{bibliography}

\end{document}